\documentclass[twocolumn,superscriptaddress,nofootinbib,showpacs,preprintnumbers]{revtex4-1}

\usepackage[latin9]{inputenc}
\usepackage{units}
\usepackage{amsmath}
\usepackage{graphicx}
\usepackage{amssymb}
\usepackage{amsfonts}
\usepackage{verbatim}
\usepackage{natbib}
\usepackage[english]{babel}
\usepackage{url}

\newcommand{\Journal}[4]{{#1} {\bf #2}, #3 (#4)}
\newcommand{\NIM}{{\em Nucl. Instrum. Methods}}
\newcommand{\NIMA}{{{\em Nucl. Instrum. Methods} A} }
\newcommand{\PLB}{{{\em Phys. Lett.} B}}
\newcommand{\RPP}{{\em Rep. Prog. Phys.}}
\newcommand{\PRP}{{{\em Phys. Rept.}}}
\newcommand{\ARNPS}{{\em Annu. Rev. Nucl. Sci.}}
\newcommand{\PR}{{\em Phys. Rev.}}
\newcommand{\APP}{{\em Astroparticle Physics}}

\newcommand{\JAP}{{\em J. Appl. Phys.}}

\newcommand{\bnel}{\mbox{$\bar{\nu}_e$} }
\newcommand{\bnmu}{\mbox{$\bar{\nu}_{\mu}$} }

\newcommand{\nel}{\mbox{$\nu_e$} }

\newcommand{\nx}{\mbox{$\nu_x$} }

\newcommand{\bildb}[7]{
\begin{figure}[#7]
    \center
    \includegraphics[width=#2]{#1}
    \parbox[t]{#4}{\caption[#6]{#3}\label{#5}}
\end{figure}
} 

\begin{document}

\title{Measurement of the proton light response of various LAB based scintillators and its implication for supernova neutrino detection via neutrino--proton scattering}

\author{B.~von~Krosigk}
\email{\url{belina.von_krosigk@tu-dresden.de}}
\affiliation{Technische Universit\"at Dresden, Institut f\"ur Kern-- und Teilchenphysik, D--01069 Dresden, Germany}
\author{L. Neumann}
\affiliation{Technische Universit\"at Dresden, Institut f\"ur Kern-- und Teilchenphysik, D--01069 Dresden, Germany}
\affiliation{now at Karlsruher Institut f\"ur Technologie, Institut f\"ur Experimentelle Kernphysik, D--76131 Karlsruhe, Germany}
\author{R. Nolte}
\affiliation{Physikalisch--Technische Bundesanstalt (PTB), Bundesallee 100, D--38116 Braunschweig, Germany}
\author{S. R\"ottger}
\affiliation{Physikalisch--Technische Bundesanstalt (PTB), Bundesallee 100, D--38116 Braunschweig, Germany}
\author{K. Zuber}
\affiliation{Technische Universit\"at Dresden, Institut f\"ur Kern-- und Teilchenphysik, D--01069 Dresden, Germany}


\begin{abstract}
The proton light output function in electron--equivalent energy of various scintillators based on linear alkylbenzene (LAB) has been measured in the energy 
range from 1\,MeV to 17.15\,MeV for the first time. The measurement was performed at the Physikalisch--Technische Bundesanstalt (PTB) using a neutron beam with continuous energy distribution. 
The proton light output data is extracted from proton recoil spectra originating from neutron--proton scattering in the scintillator. The functional behavior of the proton light output is described succesfully by Birks' law 
with a Birks constant $kB$ between (0.0094~$\pm$~0.0002)\,cm\,MeV$^{-1}$ and (0.0098~$\pm$~0.0003)\,cm\,MeV$^{-1}$ for the different LAB solutions. The constant $C$, parameterizing the quadratic term in the generalized Birks law, is consistent with zero for all investigated scintillators with an upper limit (95\% CL) of about $10^{-7}$\,cm$^2$ MeV$^{-2}$. The resulting quenching factors are especially important for future planned supernova neutrino detection
based on the elastic scattering of neutrinos on protons. The impact of proton quenching on the supernova event yield from neutrino--proton scattering is discussed.

\end{abstract}
\pacs{29.40.Mc,33.50.Hv,97.60.Bw,95.85.Ry}

\maketitle

\section{Introduction}
\label{intro}
In the last decades enormous progress has occured in the field of neutrino physics. A new field of astrophysics has been opened with the observation of solar and supernova neutrinos \cite{sno01,hir88,bio87,ale88}. First indications of ultra--high energy neutrino detection exist \cite{ish12}. From all sources the appearance of supernova neutrinos is the least predictable but a lot can be learned if such an event occurs in the Milky Way.
In the final stages of a core--collapse of massive stars nearly all the binding energy of the star is expected to be released in the form of neutrinos and antineutrinos of all flavors. The expected spectrum consists basically of two components: The deleptonization burst consisting of \nel released within a few ms and the emission of all kinds of neutrinos from the Helmholtz--Kelvin cooling phase of the protoneutron
star lasting several seconds. For recent reviews see \cite{kot06,jan07,jan12}. The intense neutrino pulse can be observed on the Earth by neutrino detectors with a sufficiently low energy threshold of about 1\,MeV. For the SN~1987A this was achieved by Kamiokande~II, IMB and the Baksan Scintillator Telescope \cite{hir88,bio87,ale88}. The signal is expected to be dominated by  \bnel because of the large cross section for the respective reaction. However, it is of great interest to measure the flux of  \mbox{$\nu_{\mu}$}, \mbox{$\nu_{\tau}$}, \bnmu and \mbox{$\bar{\nu}_{\tau}$}, collectively called \mbox{$\nu_x$}, and \nel as well. The detailed shape of the overall spectrum depends strongly on the model simulation. Recent progress has been made in using 3--dimensional simulations taking into account large scale convection. Furthermore magnetic fields, general relativity and other effects were included resulting in different spectral details. As the average neutrino energy is below about 30\,MeV, charged current interactions are only possible for electron type neutrinos, leaving only the far more challenging neutral current reactions as possible experimental detection channel for supernova $\nu_x$. In the references \cite{bea02,bea11}, neutrino--proton elastic scattering, $\nu + p \rightarrow \nu + p$, in liquid scintillator detectors is proposed for the measurement of \nx  since it is the only neutral current channel providing spectral information. Though the total cross section of this process is about a factor of four smaller than the cross section of inverse--beta decay, the yield above a realistic threshold of about 200\,keV is of the same order since this reaction is  possible for all six neutrino types \cite{bea02}.

This detection channel plays an essential role for large scale deep underground liquid scintillator detectors like KamLAND, Borexino and SNO+, in the pending measurement of the total energy and temperature of \mbox{$\nu_{\mu}$}, \mbox{$\nu_{\tau}$}, \bnmu and \mbox{$\bar{\nu}_{\tau}$}. It should be noted, however, that only a fraction of the deposited kinetic energy of highly ionizing particles like protons is visible. Hence to properly reconstruct the true kinetic proton energy $E_p$ from the visible energy $E_p^{\mathrm{vis}}$, the light response of the liquid scintillator for these particles has to be known. The ionization quenching of the light yield is well--described by Birks' law \cite{bir64} and can be quantified in terms of the parameter product $kB$ known as Birks' constant and a second parameter $C$ (Eq. (\ref{equ:birks})). While the proton light response function of the KamLAND scintillator, which consists mainly of  normal--paraffine (80\%) and pseudocumene (20\%), is already measured up to 10\,MeV \cite{bra10}, the response to protons of scintillators based on linear alkylbenzene (LAB) has not been measured before. LAB is employed for novel liquid scintillators which are used or planned as the neutrino target in recently commissioned, upcoming and potential future neutrino experiments like Daya Bay, RENO, SNO+ and LENA. Also further potential experiments like LENS, HANOHANO and Daya Bay II are considering LAB as possible solvent for their scintillator.

This article presents the measurement of the light response function $L (E)$ for protons relative to the electron response function $L_e (E)$, carried out at the accelerator facility PIAF of the Physikalisch--Technische Bundesanstalt (PTB) for different LAB based scintillators. Recoil protons are produced in the liquid scintillator by neutron--proton scattering events using a neutron beam with continuous energy distribution. The maximum energy of the recoil protons is determined from the energy of the incident neutrons, which is measured using the time--of--flight (TOF) method. Hence, the proton light output function can be determined from the position of the recoil proton edge in the pulse--height spectra produced by mono--energetic neutrons. The presented method was already used successfully at the PTB for the characterization of NE213\footnote{Nuclear Enterprise Ltd.}, BC501\footnote{Saint--Gobain Ceramics \& Plastics, Inc.} and BC501A scintillation detectors \cite{oeh08, ptb05, nov97, sch02}.

\section{Measurements}
\label{sec:meas}

In the present experiment, the light output function of binary and ternary LAB based scintillation systems is determined. In these solutions, 2,5-diphenyloxazole (PPO) is employed as primary fluor and p--bis--(o--\-methylstyryl)--benzene (bis--MSB) as secondary fluor, which acts as wavelength shifter. Two concentrations of PPO in the LAB solvent, 2\,g/l and 3\,g/l, are studied, each with and without bis--MSB, for which a concentration of 15\,mg/l is chosen. 

After dissolving the solute in LAB, the solutions were purified from oxygen by bubbling with argon for 30\,min. Without further contact to air, the solutions are subsequently filled in a scintillation detector consisting of a cylindrical dural cell with one port covered by a window made of ground and polished fused silica for higher UV transparency. The diameter and inner height of the cell is 50.8\,mm. The cell is coupled to a XP2020Q\footnote{PHOTONIS} photomultiplier tube (PMT), also equipped with a fused silica window, by a conical UV transparent poly(methyl methacrylate) (PMMA) light guide\footnote{Evonik R\"ohm GmbH }. The inner walls of the cell and the light guide are coated with BC-622A\footnote{Saint--Gobain Ceramics \& Plastics, Inc.} reflective paint to increase the light collection efficiency. The detector is equipped with a LED gain stabilization system which regulates the high voltages such that the maximum drift of the gain is smaller than 1\%. Standard NIM modules are used to obtain a pulse--height (PH) signal from the ninth dynode out of twelve, as well as a pulse--shape (PS) signal and a time--of--flight (TOF) signal from the anode of the photomultiplier. The ninth dynode is chosen for obtaining the PH signal to avoid a non--linear gain of the PMT. The PH, PS and TOF signals are registered by a multi--parameter data acquisition system and stored event--by--event. The PS discrimination module, as described in detail in \cite{mai74}, uses the zero crossing method to derive a timing marker from the trailing edge of the anode pulse. The time difference of this PS timing marker to the TOF timing marker derived from the fast leading edge of the anode pulse, using a constant fraction discriminator (CFD), is measured with a time--to--amplitude converter (TAC). It is related to the decay time of the scintillation light and therefore to the ionization density of the charged particle producing the event. The non--linearity of the TAC converter is measured using a time calibration module which produces a set of pulses at multiple intervals of $\Delta t=$~40\,ns. The variations of the peak positions from the nominal values quantify the non--linearity and are determined to be smaller than $\pm$~0.23\,ns. Fig.~\ref{fig:psd} shows the registered events sorted into a PS versus PH matrix independent of the energy of the incident neutron. The structures due to neutron--induced events ('n'), like recoil protons or other light charged particles, are partially separated from those caused by Compton electrons ('$\gamma$'). The electron events are due to ambient $\gamma$--radiation and photons resulting from the inelastic neutron scattering on carbon nuclei in the scintillator and on alumnium nuclei of the walls. A suppression of events with a PS signal below the indicated line strongly reduces the influence of photon--induced events with respect to the PH spectra of neutron--induced events. However, for LAB based scintillators a fraction of misidentified photon--induced events remains in the overlap region, as indicated by Fig.~\ref{fig:psd}. This is mainly effected by the small number of photoelectrons produced per scintillation event which is primarily due to the wavelength mismatch between the emission spectrum and the sensitivity of the photocathode.

\bildb{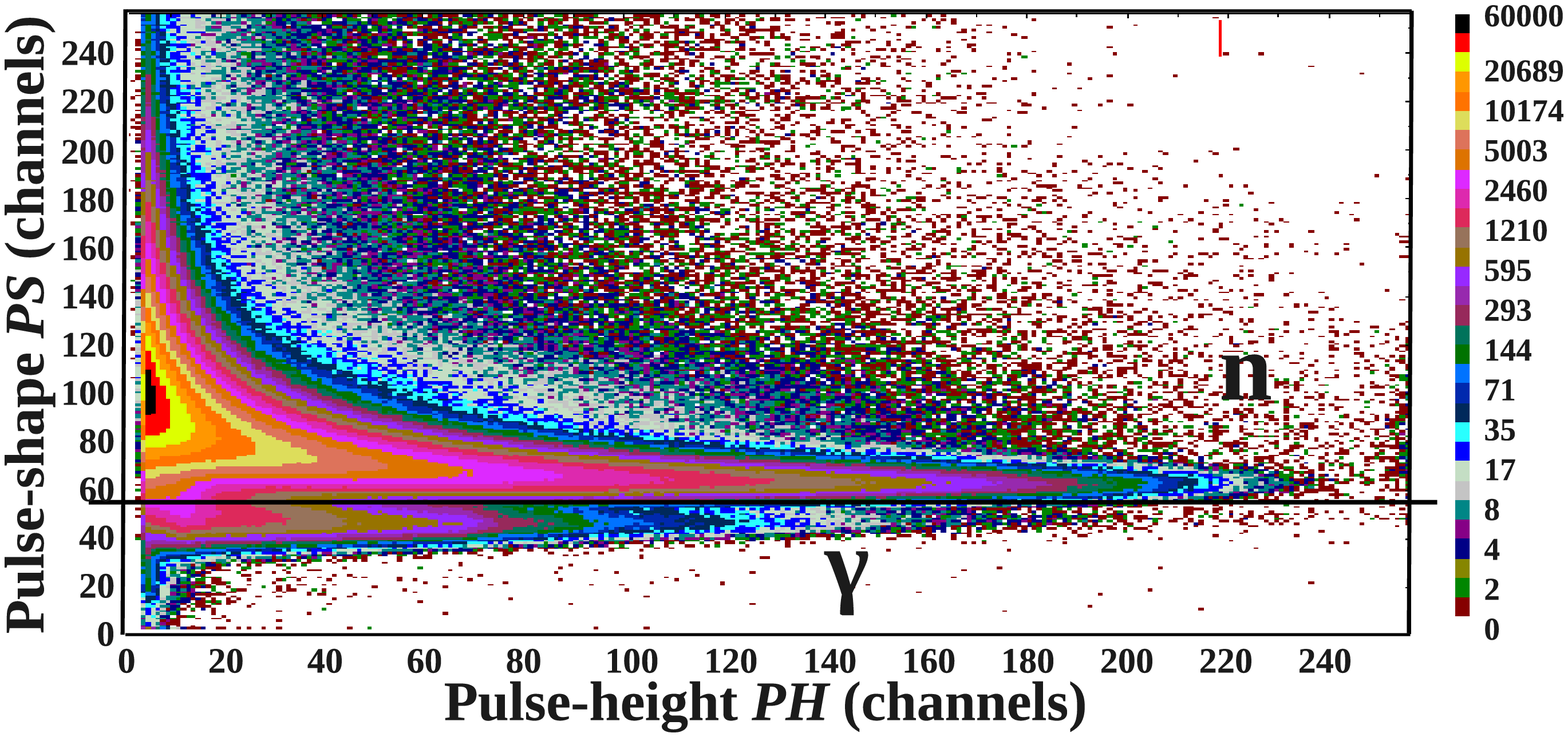}{0.49\textwidth}{Detected events sorted into a pulse--shape (PS) vs. pulse--height (PH) matrix in low gain (LG) mode and with selection of the energy of the incident neutrons. The color scale defines the intensity in counts. The structures labeled with 'n' and '$\gamma$' are caused by charged particles resulting from neutron interactions with the scintillator and by Compton electrons from photon interactions respectively. The photons result partly from ambient $\gamma$--radiation and partly from inelastic neutron scattering. $n/\gamma$--separation, indicated by the solid line, can not fully separate neutron-- and photon--induced events due to the small number of photoelectrons produced per scintillation light photon. The scintillator contained LAB + 2\,g/l PPO + 15\,mg/l bis--MSB.}{0.48\textwidth}{fig:psd}{}{t}

The energy calibration of the PH spectra and determination of the light response to electrons for each scintillator is obtained from the position of the Compton edges produced by a set of standard $\gamma$--ray sources ($^{137}$Cs, $^{22}$Na, $^{207}$Bi). Ambient background is measured separately and subtracted from each source measurement. The analysis of the calibration data is described in detail in Sec.~\ref{sec:ana}.

To obtain a proton light output function over an energy range as large as possible in a single experiment, a neutron beam with a continuous spectral distribution is used, which is produced at the PTB TOF facility by bombarding a stopping--length beryllium target with a 19\,MeV proton beam from the isochronous CV28 cyclotron \cite{bre89}. At an emission angle of $0^{\circ}$, the maximum energy of the neutrons is 17.15\,MeV. The center of the cell was positioned at a distance of 12.11\,m from the Be target under $0^{\circ}$ with respect to the proton beam. The distance between target and detector cell is measured with an accuracy of $\pm$~2\,mm. To consider the target thickness of 3\,mm and include the uncertainty of the effective center of the cell caused by the fluence attenuation, $\pm$~6\,mm are conservatively assumed. With a beam current of about 50\,nA, the neutron rate at the detector front face was about $3\times10^4$\,s$^{-1}$. The repetition frequency of the proton beam was set to 481.3\,kHz. The trigger rate of the CFD module was about $3\times10^3$\,s$^{-1}$ for a \mbox{threshold} set close to the electronic noise. The probability for multiple neutron interactions within one beam pulse is thus smaller than 1\% and does not cause a deterioration of the shape of the pulse--height spectra. In the given configuration the neutron energy threshold for time--frame overlap is 0.15\,MeV, which corresponds to an electron--equivalent energy for the given scintillators of less 
than 20\,keV. Hence, with a hardware threshold of about 200\,keV, time--frame overlap is avoided.

Proton bunches that are not entirely deflected by the internal beam pulse selector system cause additional pulses, so--called satellite pulses, on the target. The spectral distributions of outgoing satellite neutrons are assumed to be the same as the ones from the main proton bunches and are visible in the PH vs. TOF matrix as faint distributions which are shifted in TOF relative to the distribution resulting from the main beam pulse. Hence, satellites cause spurious events above the recoil proton edge in the pulse--height spectra which could affect the determination of the position of the recoil proton edge. A subtraction of the shifted TOF spectra, though, showed no significant influence on the position determination of the proton recoil edge.

To improve the dynamic range of the pulse--height measurement for neutron energies below 5\,MeV, each measurement is repeated with an increased amplification. This amplification mode is referred to as high gain (HG) mode, the other one as low gain (LG) mode. For the HG measurement, $n/\gamma$--separation by pulse--shape discrimination (PSD) is not possible. However, at these energies, inelastic scattering on carbon does not contribute to the neutron induced response because of the energy of the first excited state in $^{12}$C which is at 4.439\,MeV. The contribution from prompt $\gamma$--radiation from the target and from ambient $\gamma$--radiation is subtracted experimentally.

\section{Analysis}
\label{sec:ana}

\subsubsection{Calibration}
\label{sssec:cal}
The amount of detected scintillation light and the pulse--height resolution strongly depend on the specific composition of the scintillator. Therefore the pulse--height calibration of the scintillation detector is repeated for each filling and amplification mode using three $\gamma$--ray sources with a total of six $\gamma$--lines, namely $^{137}$Cs ($E_\gamma=0.662$\,MeV), $^{22}$Na ($E_\gamma=0.511$\,MeV, $E_\gamma=1.275$\,MeV) and $^{207}$Bi ($E_\gamma=0.570$\,MeV, \mbox{$E_\gamma=1.064$\,MeV}, $E_\gamma=1.770$\,MeV).

The position of each Compton edge, with energy

\begin{equation}
\label{equ:compton}
E_e  = \frac{2E^2_{\gamma}}{0.511 + 2E_{\gamma}},
\end{equation}

where all energies are given in MeV, is defined by comparing the measured pulse--height spectrum with a distribution simulated using the Monte Carlo photon transport code \textsc{gresp} \cite{die82}. To adjust the theoretical shape to the measurement, the simulated spectrum is folded using a Gaussian resolution function with a pulse--height dependent width and fitted to the measured pulse--height spectra (Fig.~\ref{fig:215HGCs}).

\bildb{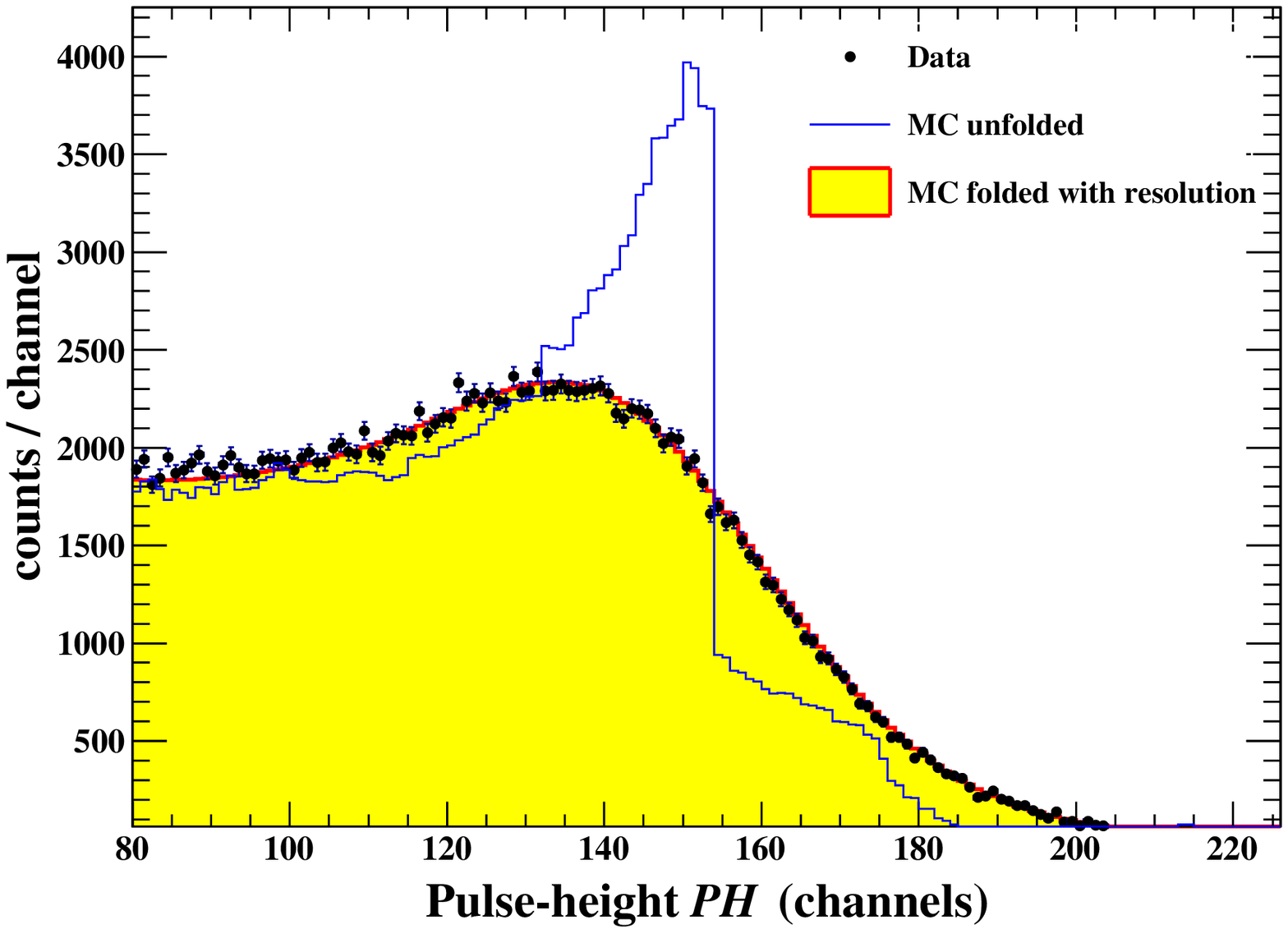}{0.5\textwidth}{Pulse--height spectrum of $^{137}$Cs given in pulse--height channels. The calculated Compton electron spectrum before and after folding with the Gaussian resolution function is compared to the experimental spectrum for LAB with 2\,g/l~PPO and 15\,mg/l~bis--MSB, measured in HG mode.}{0.48\textwidth}{fig:215HGCs}{}{h}

In this way the measured pulse--height signal PH is calibrated in units of the maximum kinetic energy $E_e$ of the Compton electrons. The calibrated pulse--height (light output) $L$ is given by

\begin{equation}
\label{equ:conv}
L = n \cdot PH,
\end{equation}

where $n$ denotes the calibration factor. The relative resolution $\Delta L$ at pulse--height $L$ is parameterized by \cite{sch80}

\begin{equation}
\label{equ:resol}
\frac{\Delta L}{L} = \sqrt{\alpha^2 + \frac{\beta^2}{L} + \frac{\gamma^2}{L^2}}.
\end{equation}

The fit parameters $\alpha$, $\beta$, $\gamma$ can be attributed to the individual contributions from spatial dependence of the light collection efficiency, statistical variation of the number of photoelectrons  and electronic noise, respectively. The values of the resolution parameters for each scintillator filling are obtained by fitting Eq.~(\ref{equ:resol}) to the locally determined full width at half maximum $\Delta L$ at each Compton edge position. Fig.~\ref{fig:PHresol} shows the resulting resolution functions for every sample as a function of the light output. The lower resolution for the scintillators without wavelength shifter is mainly attributed to detector properties. The reflectivity of the BC--622A coating inside the cell is reduced below 420\,nm and the XP2020Q PMT has its peak sensitivity at 420\,nm, while the emission peak of PPO lies around 360\,nm. Thus the spatial dependency and statistical variation are increased. However, the properties are more fortunate for the scintillators with bis--MSB which has a maximum emission around 420\,nm. For this reason, besides the two binary scintillators, the two systems with additional wavelength shifter are investigated, though no difference in the ionization quenching properties is to be expected.

\bildb{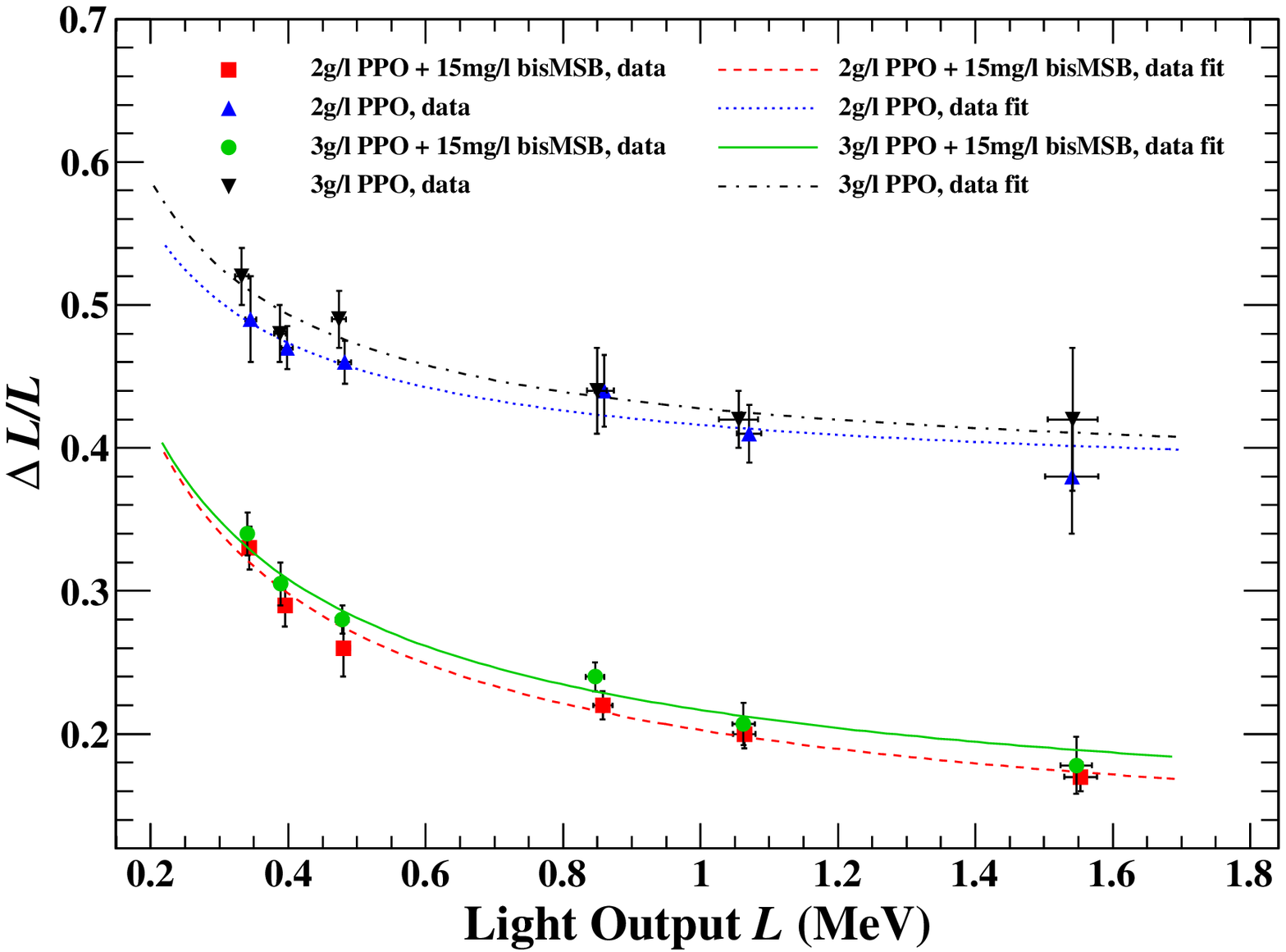}{0.5\textwidth}{Relative pulse--height resolution $\Delta L/L$ for electrons as function of the light output $L$ in electron--equivalent energy.  The respective solute admixture to LAB is given in the legend. Shown are the data points 
acquired from source measurements in HG mode and the respective adapted resolution 
function Eq.~(\ref{equ:resol}). In the shown total uncertainties, the single contributions are added quadratically.}{0.48\textwidth}{fig:PHresol}{}{h}

According to \cite{bir64}, the light output function $L(E)$ is related to the stopping power $dE/dx$ of a charged particle of kinetic energy $E$ in the traversed scintillator via

\begin{equation}
\label{equ:orig_birks}
L(E) = S \cdot \int_{0}^{E} dE \left[1+kB\left(\frac{dE}{dx}\right)\right]^{-1},
\end{equation}

if the particle is stopped in the scintillator. This relation is known as Birks' formula with the product $kB$ being the Birks constant. The light output $L$ is the total light emitted when a charged particle loses all of its energy $E$ within the scintillator. For electrons with energies \mbox{$\geq 125$\,keV}, the stopping power becomes very small \cite{bir64} and Eq.~(\ref{equ:orig_birks}) reduces to the linear expression

\begin{equation}
\label{equ:Le}
L_e(E) \approx S \cdot (E - E_0).
\end{equation}

The subscript '$e$' is added here to underline that this description of the light output only holds for electrons. The parameter $S$ denotes the chosen scaling between electron energy and light output and is set to $S=1$, which yields the light output in electron--equivalent energy, if the calibration procedure described above is used. The constant $E_0$ accounts for the fact that the response of the scintillator to electrons is only approximately linear. The non--linearity at small energies leads to an energy offset, which is assumed to be $E_0=5$\,keV for NE213 and BC501. Recent measurements with LAB and \mbox{EJ--301} scintillators \cite{wan11}, though, revealed a non--linear behavior already starting at 400\,keV, resulting in an offset of about 50\,keV. Within the present work, the potential non--linearity of the electron light output can not be discriminated from the effect of the remaining electronic offset of the amplifiers which was corrected for during the analysis of the pulse--height calibration, i.e. $E_0$ is set to zero.

A comparison of Eq.~(\ref{equ:conv}) and Eq.~(\ref{equ:Le}) gives the relation between PH and 
electron energy
\begin{eqnarray}
\label{equ:eresp}
PH &=& n^{-1} \cdot (E - E_0) \nonumber\\
&=& n^{-1} \cdot E,
\end{eqnarray}
where the second equality only holds for the case described here. The calibration factor $n^{-1}$ can now be derived from a linear fit to the positions of all 
Compton edges in the PH spectrum as performed in Fig.~\ref{fig:eresp_abs} and is independent of the total offset of the pulse--height scale. The values obtained in the present measurements are summarized in Tab.~\ref{tab:yield}.

\bildb{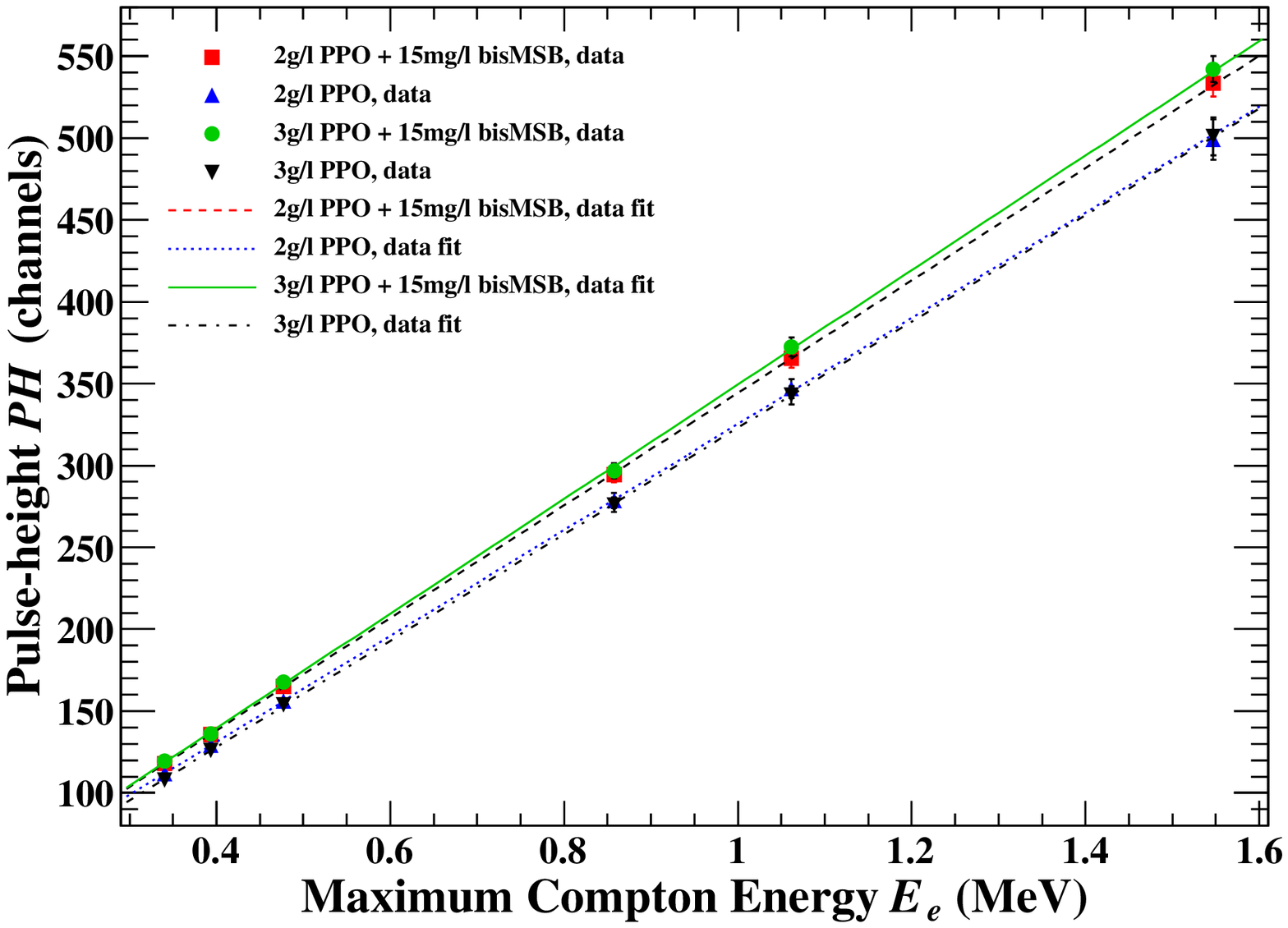}{0.5\textwidth}{Position of the Compton edges in pulse--height channels as function of the maximum kinetic energy $E_e$ of the Compton electrons. The data points are taken in HG mode. The linear fits describe the response of different LAB based scintillators to electrons. The respective solute admixture is given in the legend. In the shown total uncertainties, the single contributions are added quadratically.}{0.48\textwidth}{fig:eresp_abs}{}{hbtp}

\setlength{\tabcolsep}{12pt}
\renewcommand{\arraystretch}{1.5}
\begin{table}[htbp]
\begin{center}
\caption{\label{tab:yield} Calibration factor $n^{-1}$ of the investigated LAB based 
scintillators obtained in LG and HG mode. The given 1$\sigma$ uncertainties are the results of a least--squares fit to the data points shown in Fig.~\ref{fig:eresp_abs}.}
\begin{tabular*}{0.48\textwidth}{lll}
\hline\noalign{\smallskip}
LAB admixture & Gain & $n^{-1}$ \\
 &  & [MeV$^{-1}$] \\
\noalign{\smallskip}\hline\noalign{\smallskip}
2g/l PPO, 15mg/l bis-MSB & LG & 324.5 $\pm$ 1.0\\
 & HG & 343.4 $\pm$ 1.1\\
2g/l PPO & LG & 297.8 $\pm$ 1.4\\
 & HG & 324.0 $\pm$ 2.1\\
3g/l PPO, 15mg/l bis-MSB & LG & 325.5 $\pm$ 0.9\\
 & HG & 350.5 $\pm$ 0.7\\
3g/l PPO & LG & 294.7 $\pm$ 2.0\\
 & HG & 325.1 $\pm$ 2.5\\
\noalign{\smallskip}\hline
\end{tabular*}
\end{center}
\end{table}

\subsubsection{Proton light output function}
\label{sssec:Lp}
For the determination of the proton light output function $L(E)$, basically the same technique is used as for the analysis of the photon--induced events in the calibration measurements described in Sec.~\ref{sssec:cal}. The events measured with the white neutron beam are first sorted into a TOF vs. PH matrix. As mentioned above, PSD is used to select only events produced by light charged particles for the spectra obtained with the LG setting of the amplifier, while in the HG mode $n/\gamma$--separation by PSD is not feasible. The TOF range covers the full neutron energy range from about 1\,MeV to 17.15\,MeV. The TOF scale is established using the arrival time difference between prompt gammas and neutrons of maximum energy and a time determined using an electronic time calibration module. The prompt gamma line is broadened due to the finite time resolution of the accelerator and the uncertainty of the centroid position of the resulting peak is calculated to be $\pm$~0.1\,ns. Calibration gives a TOF channel width of 0.7764\,ns and the relative uncertainty of the time calibration is 0.05\%. Pulse--height spectra induced by quasi mono--energetic neutrons are then extracted by collecting all events contained in small windows around the TOF value corresponding to an individual neutron energy. The size of the TOF window is always made smaller than the pulse--height resolution $\Delta L$ for protons of the respective energy. A second window of the same size is placed on the random background at TOF values smaller than those of neutrons with the maximum energy and the extracted background spectrum is subtracted from the signal spectrum. Pulse--height spectra for different incoming neutron energies are shown in Fig.~\ref{fig:PHspec}. The energy of the recoil protons, producing the edge in the pulse--height spectra, equals that of the incident neutrons. 
\bildb{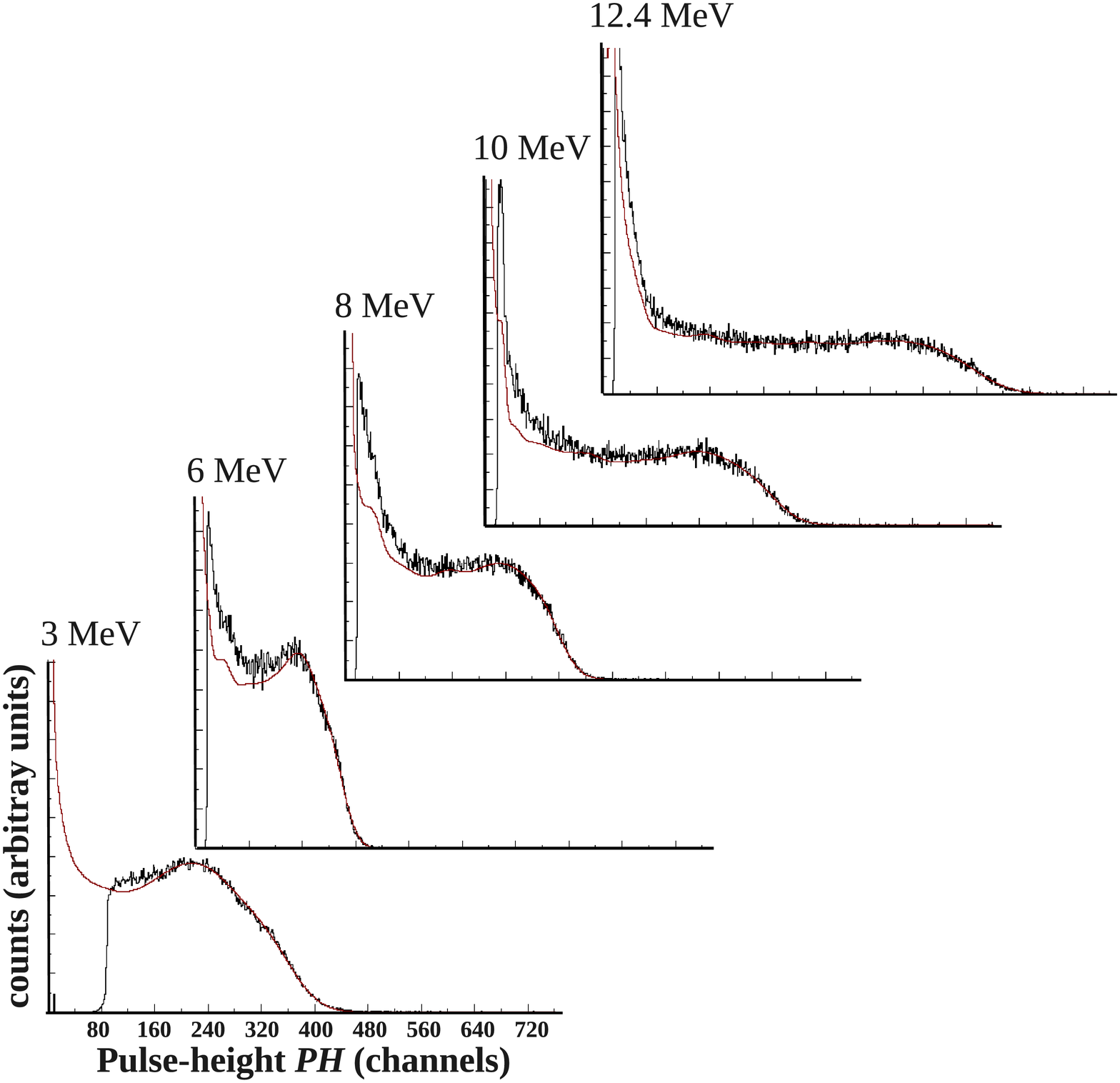}{0.48\textwidth}{Calculated (line) and measured (histogram) proton recoil spectra at different incoming neutron energies. The measured spectra are extracted from the data sets for LAB with 2\,g/l~PPO and \mbox{15\,mg/l~bis--MSB}. For the spectrum at 3\,MeV neutron energy, HG data is used, for the rest LG data. The calculation is performed using the \textsc{nresp7} code and does not consider a threshold (leading to counts starting at channel 1, where there is no data available) and photon induced background (leading to less counts at low pulse--heights above threshold compared to data).}{0.48\textwidth}{fig:PHspec}{}{t}

Before extracting the position of the recoil edge, the dependence of the TOF measurement on the pulse--height is investigated. The so--called time--walk is caused by imperfections in the constant fraction timing technique employed in the CFD module which show up in particular at small pulse--height. It leads to a distortion of the PH vs. TOF matrices and finally in the extracted proton recoil spectra at low pulse--heights. A correction of the PH vs. TOF matrices of LAB with 2\,g/l PPO and 15\,mg/l bis-MSB revealed a deviation of about only one channel in the TOF spectrum. For comparison, the smallest window width, which is chosen at high neutron energies, has a width of five channels. A spectral distortion in the region of small pulse--heights, however, has the biggest influence on the definition of the proton recoil edge at low neutron energies, where the TOF window width has to be increased because of a decreased resolution, as described above. The maximum width of the TOF windows amounts to 25 channels, which is much larger than the observed deviation in the measured TOF. Hence, time--walk has a minor effect. As a correction of the PH vs. TOF matrices, to account for time--walk, can bring in new artefacts due to the binned structure of the data, no correction is performed within this analysis.

To calculate the detector response to monoenergetic neutrons, the Monte Carlo code \textsc{nresp7} \cite{nresp} is used. \textsc{nresp7} models all reactions of neutrons in the scintillator, in the detector housing and in the Lucite light guide, including the production of secondary charged particles, but it does not simulate the interaction of deexcitation photons (e.g. from the first excited state in $^{12}$C at 4.439\,MeV) resulting from inelastic neutron scattering. Instead, it is assumed that such events are suppressed by PSD. 
If the incoming neutrons have enough energy to excite $^{12}$C, signals from deexcitation photons with pulse--heights up to about 180\,channels can add up to the proton recoil spectra. As this background is not fully suppressed by PSD, it leads to more counts above threshold in the measured pulse--height spectra than in the simulated ones (see Fig.~\ref{fig:PHspec}) for neutron energies higher than $\sim$4.5\,MeV. However, this mainly affects the proton recoil edge determination, if the recoil edge is in the same region as the photon induced background. This is only the case for a maximum proton energy (i.e. incoming neutron energy) up to $\sim$7\,MeV. 
For the fit of calculated to the measured spectrum, a pulse--height interval is chosen, which starts below the proton recoil edge and ends above. For neutron energies above about 7\,MeV, this interval is chosen such, that photon induced background is excluded.
The pulse--height given in channels is finally translated into light output in electron--equivalent energy $L$ using the factor $n^{-1}$ obtained in the calibration with photon sources. The resulting relative proton light output resolution functions are plotted in Fig.~\ref{fig:Lpresol}.

\bildb{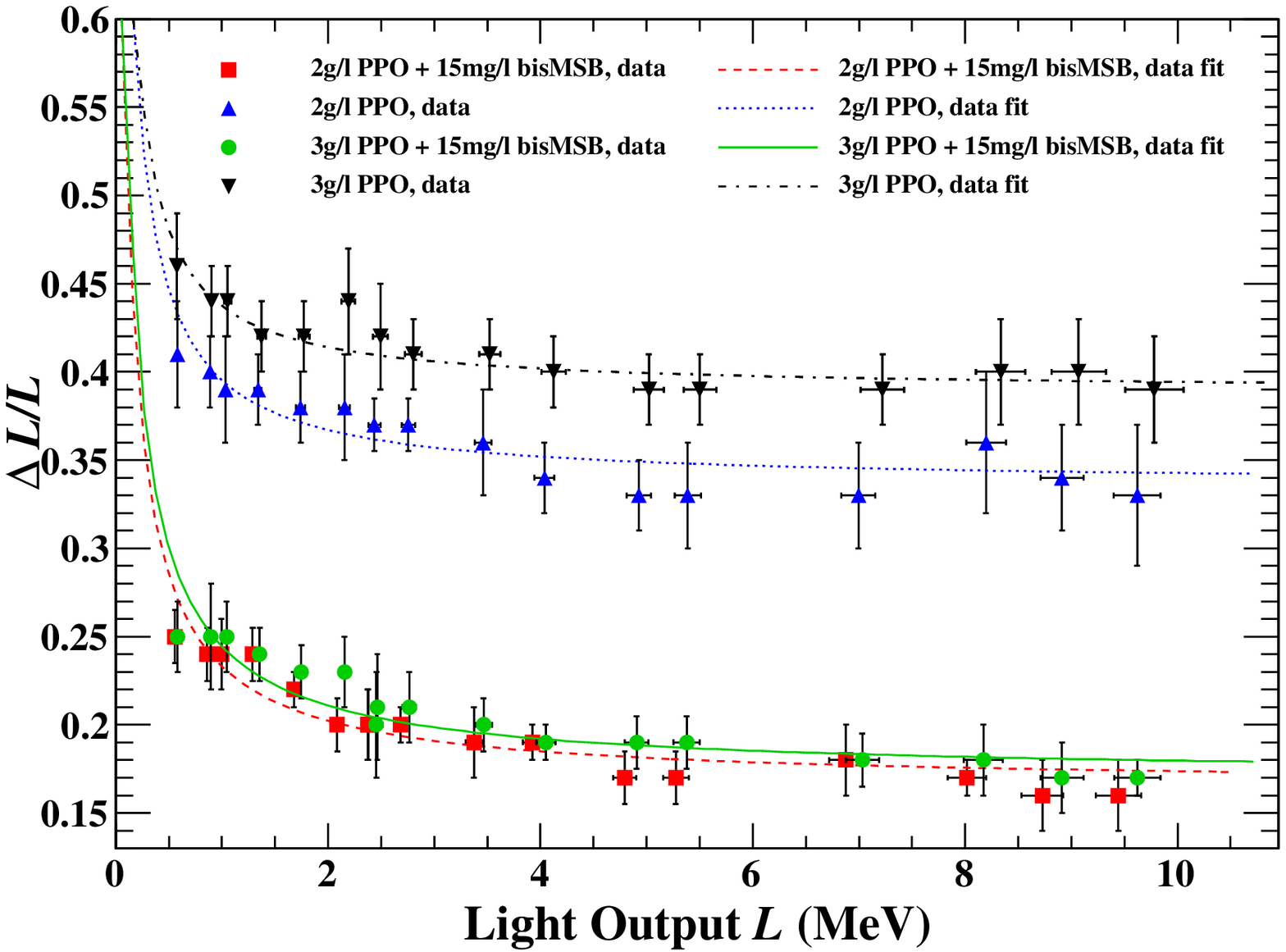}{0.5\textwidth}{Relative pulse--height resolution $\Delta L / L$ for protons as function of the light output $L$ in electron--equivalent energy for four different LAB based scintillators. The respective solute admixtures are given in the legend. In the shown total uncertainties, the single contributions are added quadratically.}{0.48\textwidth}{fig:Lpresol}{}{h}

The simulated pulse--height spectra are adapted to the experimental ones in an iterative process. As an initial step, an existing reference light output function for NE213 detectors is used to calculate pulse--height spectra using \textsc{nresp7}. The simulated spectra are compressed or stretched and adjusted in height to fit the experimental spectra in the pulse--height region around the recoil edge  and the respective compression or stretching factors are applied to correct the light output function. The corrected function is inserted as input for the next iteration step. The behavior of the applied light output function follows a straight line for $E\geq8$\,MeV, where the parameters of the line are gained from a fit to the extracted data points. After about four steps, the procedure converges. Fig.~\ref{fig:respsNLICHT} compares the initial light response function with the final one for LAB with 2\,g/l~PPO and 15\,mg/l~bis--MSB. The data points extracted in this way are presented in Fig.~\ref{fig:resps215} together with the electron data points gained from calibration. The results for the remaining scintillators are shown in Fig.~\ref{fig:resps200}~--~\ref{fig:resps300}. The contributions to the total uncertainty in $L$ and $E$ are summarized in Tab.~\ref{tab:syst}.

\bildb{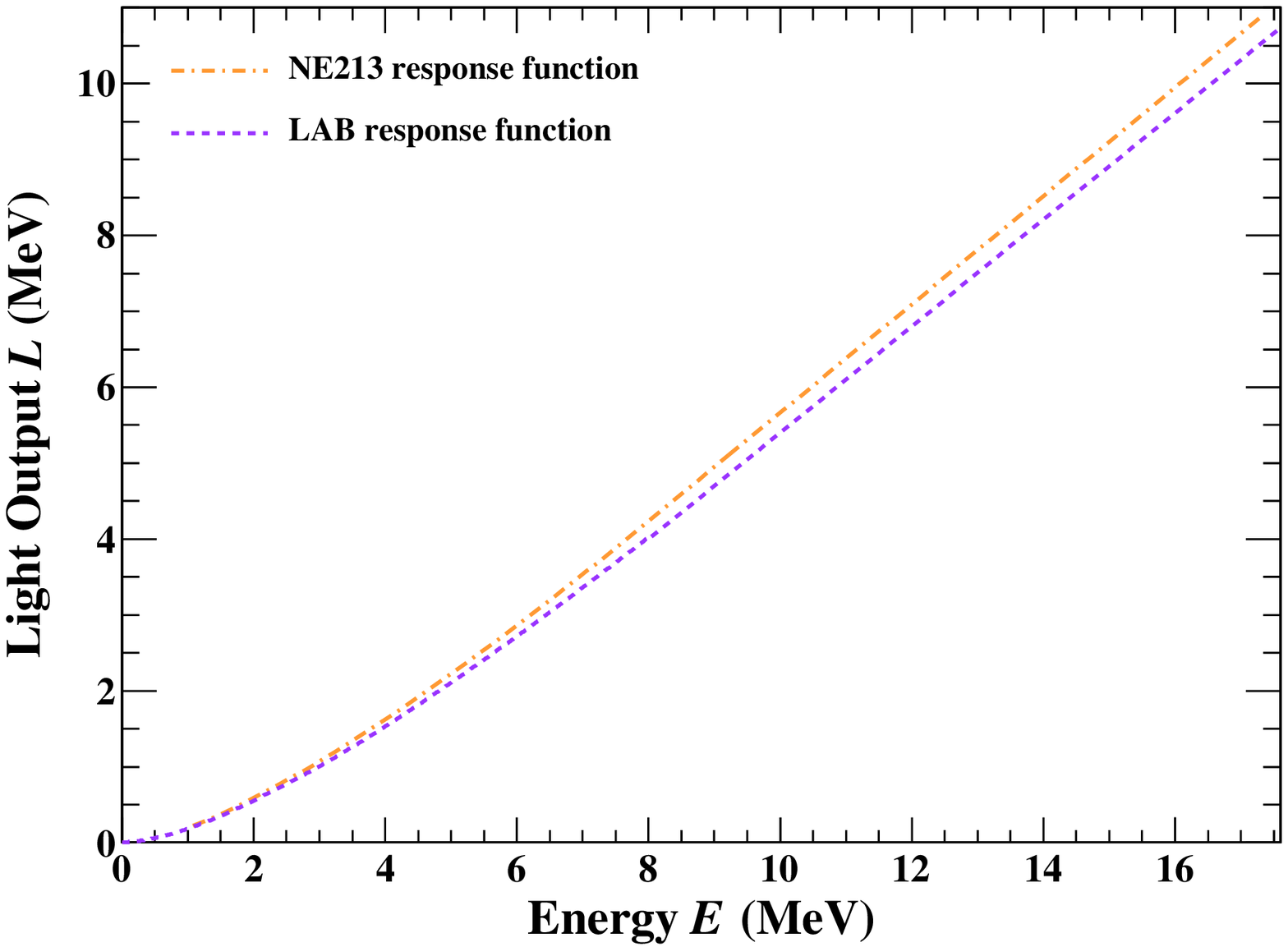}{0.5\textwidth}{Light output $L$, given in electron--equivalent energy, as function of kinetic energy $E$. Shown are the light output functions inserted in \textsc{nresp7} for the initial (dashed dotted line) and the final (dotted line) proton spectrum calculations for LAB with 2\,g/l~PPO and 15\,mg/l~bis--MSB.}{0.48\textwidth}{fig:respsNLICHT}{}{htbp}
\bildb{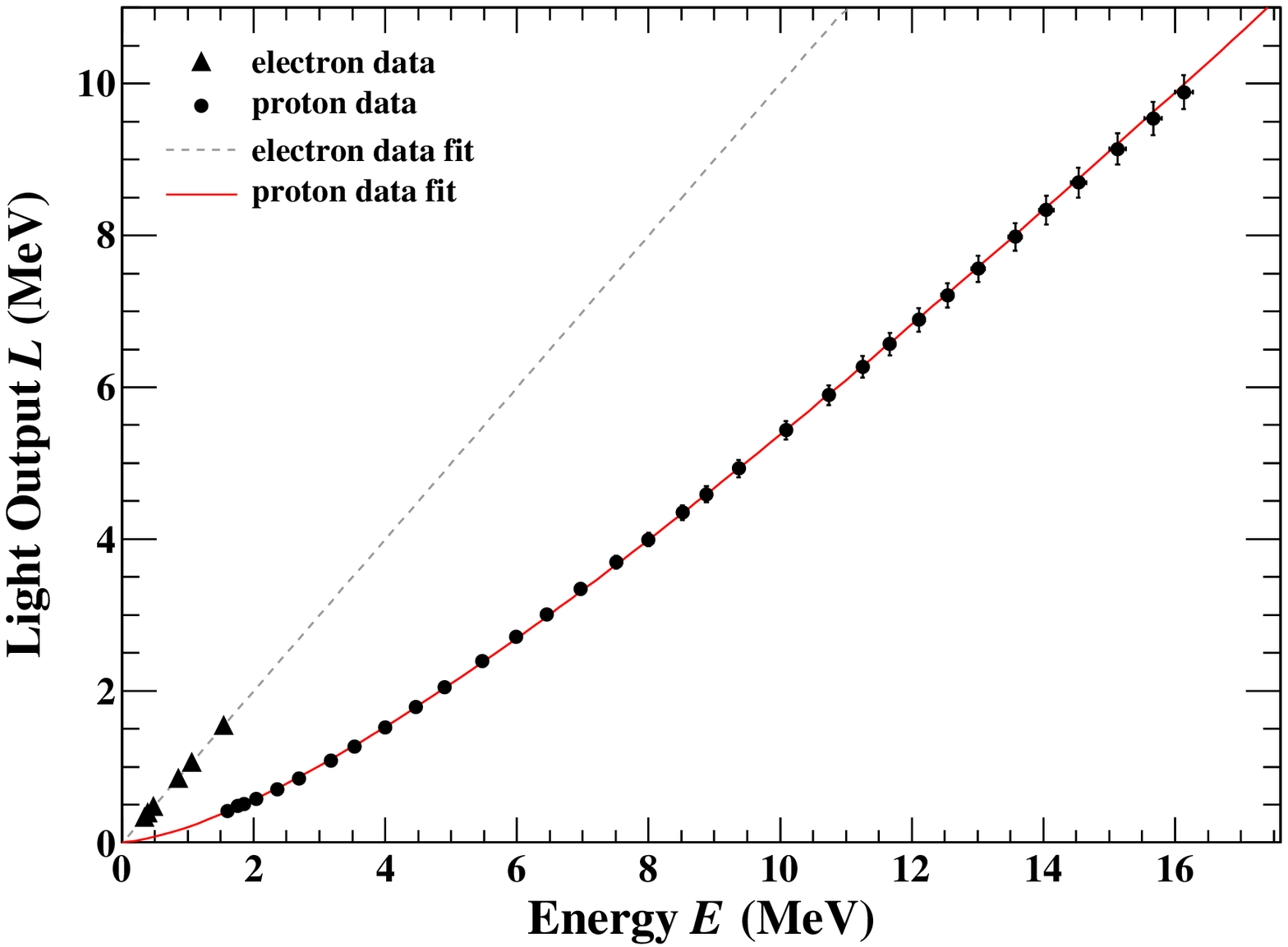}{0.5\textwidth}{Light output $L$, given in electron--equivalent energy, as function of kinetic energy $E$. Electron data is fitted with Eq.~(\ref{equ:Le}) and proton data with Eq.~(\ref{equ:birks}). The data is taken with LAB, 2\,g/l~PPO and 15\,mg/l~bis--MSB.  In the shown total uncertainties, the single contributions (listed in Tab.~\ref{tab:syst}) are added quadratically. For some values of $L(E)$, the resulting error bars are smaller than the marker size.}{0.48\textwidth}{fig:resps215}{}{htbp}
\bildb{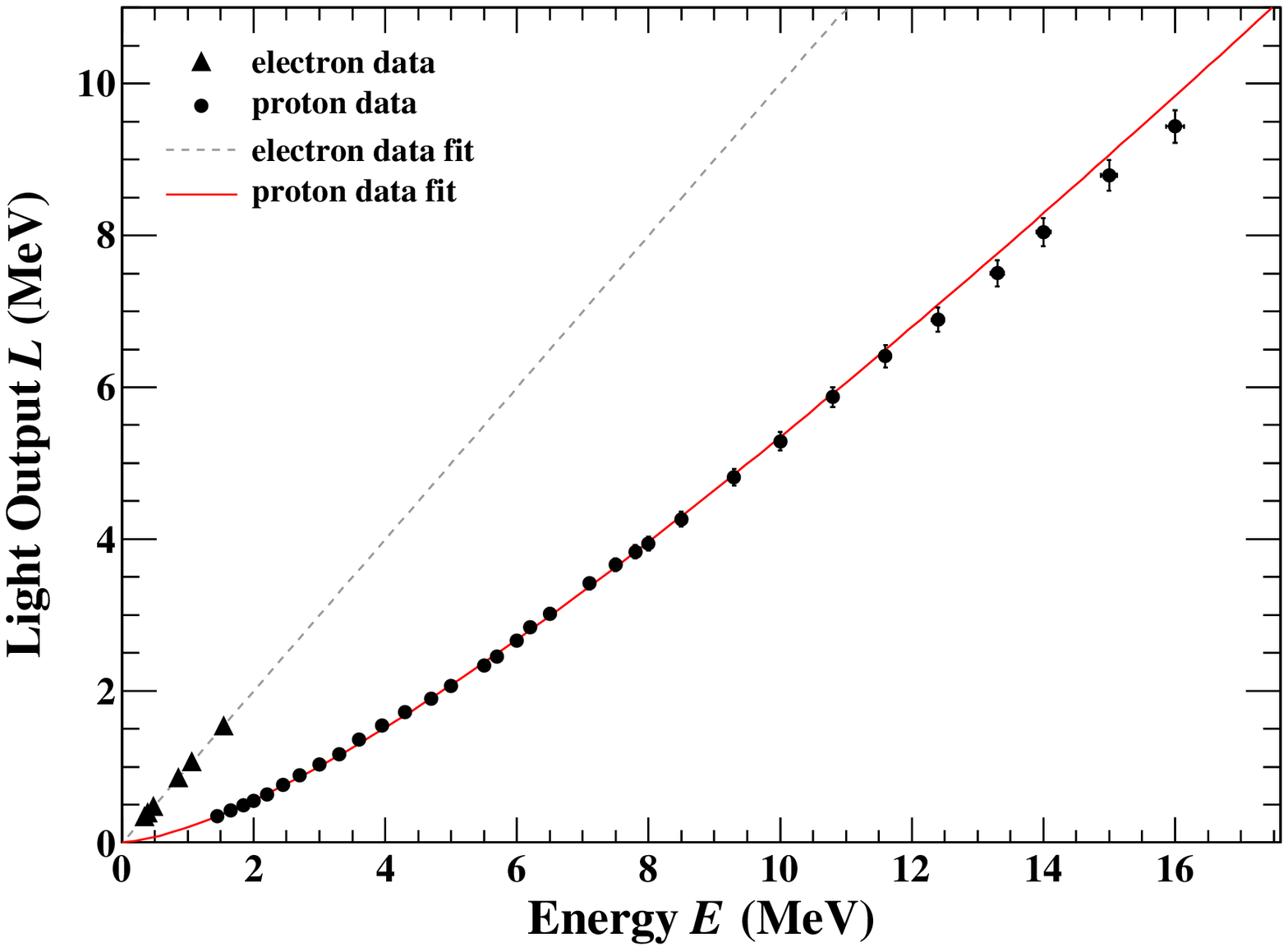}{0.5\textwidth}{Same as Fig.~\ref{fig:resps215} but for LAB with 2\,g/l~PPO, without bis--MSB.}{0.48\textwidth}{fig:resps200}{}{htbp}
\bildb{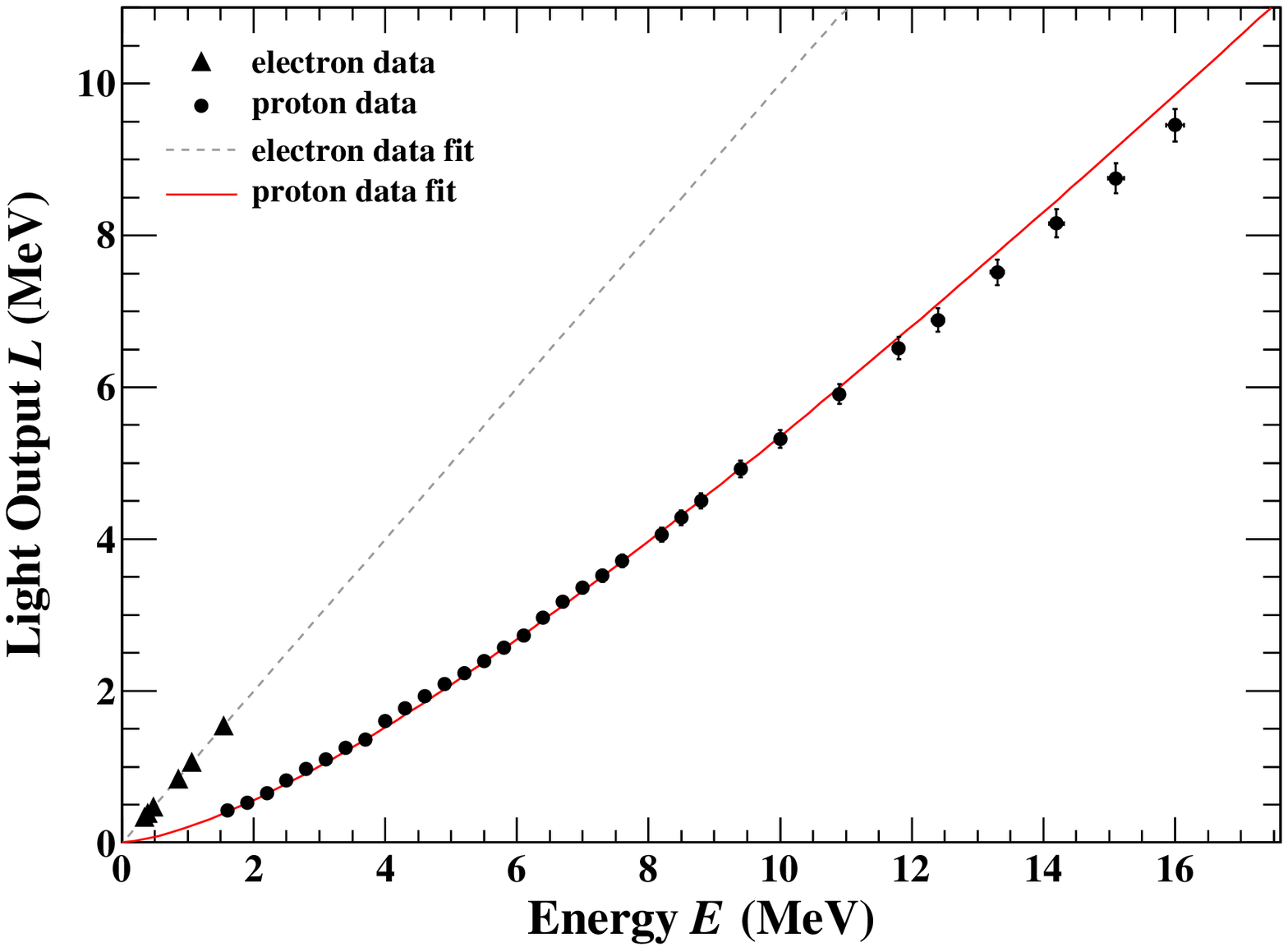}{0.5\textwidth}{Same as Fig.~\ref{fig:resps215} but for LAB with 3\,g/l~PPO and 15\,mg/l~bis--MSB.}{0.48\textwidth}{fig:resps315}{}{htbp}
\bildb{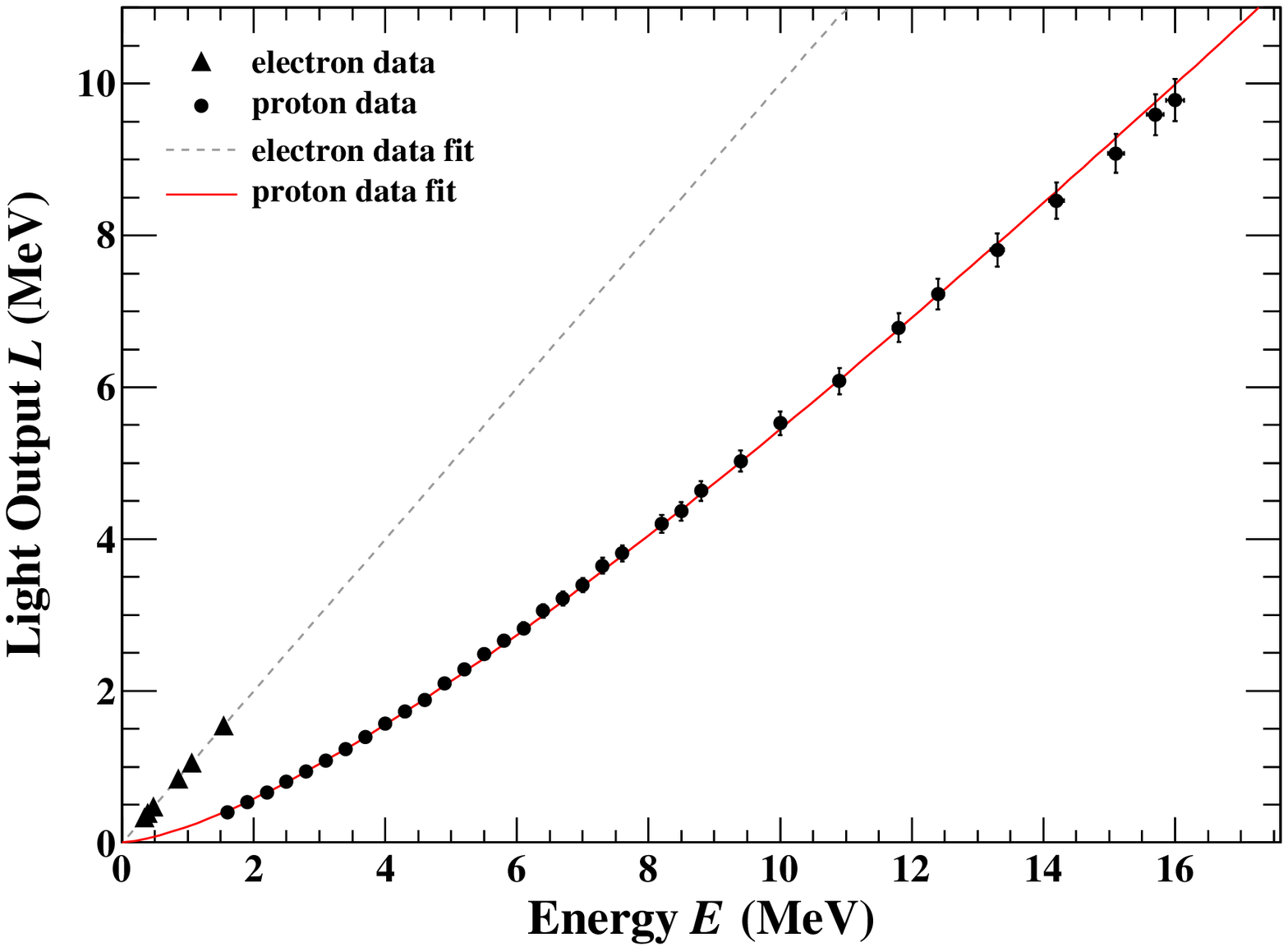}{0.5\textwidth}{Same as Fig.~\ref{fig:resps215} but for LAB with 3\,g/l~PPO, without bis--MSB.}{0.48\textwidth}{fig:resps300}{}{htbp}

\setlength{\tabcolsep}{17pt}
\begin{table}[htbp]
\begin{center}
\caption{\label{tab:syst}  Systematic and statistical uncertainties of the measurement of the relative proton light output $L(E)$ and their 1$\sigma$ values. Time--walk of the CFD, satellite pulses, multiple neutron events and time--frame overlap have a minor effect and are neglected. The extracted proton recoil edge position carries two uncertainties. The statistical uncertainty is the uncertainty on the compression factor within the adaption of calculated to measured proton recoil spectrum. Furthermore, since the choice of the intervall for the fit shifts the recoil edge position by more than 1$\sigma_{\mathrm{stat.}}$, this shift is considered as additional systematic uncertainty.}
\begin{tabular*}{0.48\textwidth}{ll}
\hline\noalign{\smallskip}
{\bf Systematic uncertainty} & 1 $\sigma_{\mathrm{syst.}}$ \\
\noalign{\smallskip}\hline\noalign{\smallskip}
prompt $\gamma$ peak centroid position & $\pm$0.1\,ns\\
TAC non--linearity & $\pm$0.23\,ns\\
time calibration & $\pm$0.05\%\\
target -- detector distance & $\pm$6\,mm \\
gain stabilization & $\pm$1\%\\
pulse--height offset & $\pm$2\,ch\\
recoil edge position & $\pm$2\%\\
calibration factor & (see Tab.~\ref{tab:yield})\\
\noalign{\smallskip}\hline
\hline\noalign{\smallskip}
{\bf Statistical uncertainty} & 1 $\sigma_{\mathrm{stat.}}$ \\
\noalign{\smallskip}\hline\noalign{\smallskip}
recoil edge position & $<$0.02\%\\
\noalign{\smallskip}\hline
\end{tabular*}
\end{center}
\end{table}
The most popular analytical description for the energy dependency of the light output is given by Birks' formula Eq.~(\ref{equ:orig_birks}), assuming ionization quenching to be the reason for the reduced light output. This expression, though widely used for all kinds of scintillators, was originally developed for anthracene crystals. However, the underlying physical processes in liquid scintillators are different and this description should be considered as being semi--empirical. For large values of $dE/dx$, such as for incident ions, Eq.~(\ref{equ:orig_birks}) is often encountered in its generalized form

\begin{equation}
\label{equ:birks}
L(E) = S \cdot \int_{0}^{E} dE \left[1+kB\left(\frac{dE}{dx}\right) + C\left(\frac{dE}{dx}\right)^2\right]^{-1},
\end{equation}

including a quadratic correction term parameterized by $C$, as proposed in \cite{cho52}. Within the present work, electron and proton light output functions of a particular scintillator are measured under identical conditions so that the scaling factor $S$ for proton data is the same as the one for electron data, i.e. $S=1$. The Birks constant $kB$ and the parameter $C$ for LAB based scintillators are obtained by means of a $\chi^2$ fit of the theoretically expected light output $L^{\mathrm{theo}}$ in Eq.~(\ref{equ:birks}) to the experimentally determined values $L^{\mathrm{exp}}$ presented in Fig.~\ref{fig:resps215}~--~\ref{fig:resps300}. Extracted data points, instead of the final input function in \textsc{nresp7}, are used for this fit to be independent from the mentioned straight line fit for \mbox{$E\geq8$\,MeV}. The stopping powers $dE/dx$ for protons are calculated using the code \textsc{srim} \cite{srim}. The $\chi^2$ calculation is carried out following the pull approach \cite{fog02}, where

\begin{eqnarray}
\label{equ:chi2}
\chi^2(kB,C,\xi _k) &=& \sum_{n=1}^N \frac{\left[L_n^{\mathrm{exp}} - L_n^{\mathrm{theo}} \left(1  + \sum_{k=1}^K \xi _k f_n^k \right)\right]^2}{u_n^2} \nonumber\\
&+& \sum_{k=1}^K \xi _k^2.
\end{eqnarray}

Within this calculation $K=8$ systematic uncertainties, listed in Tab.~\ref{tab:syst}, are included as nuisance parameters. $f^k_n$ describes the fractional change of the $n$--th value of $L^{\mathrm{theo}}$ if the $k$--th source of systematics is varied by 1\,$\sigma_k$ and $\xi_k$ is a standard normal deviate. The normalization condition for the $\xi_k$'s is realized through quadratic penalties, summed over the $K$ sources of systematics. The statistical uncertainty, denoted by $u$, is also indicated in Tab.~\ref{tab:syst}. For each combination of $kB$ and $C$, Eq.~(\ref{equ:chi2}) is minimized with respect to all $\xi_k$'s using \textsc{minuit} and the minimal value, $\chi_{\mathrm{pull}}^2$, is stored in a $(kB,C)$ map. The best fit values are finally extracted at the minimal $\chi_{\mathrm{pull}}^2$ in the full parameter space. The uncertainties are extracted from the $\chi_{\mathrm{pull}}^2$ projections on the $kB$ resp. $C$ axis. The best fit value of $kB$ has a two--sided limit and the $kB$ values at $\chi_{\mathrm{pull}}^2 -  \chi_{\mathrm{pull,min}}^2 = 1$ give the $1\sigma$ uncertainty. $C$ has a one--sided limit and the $C$ value at $\chi_{\mathrm{pull}}^2 -  \chi_{\mathrm{pull,min}}^2 = 1.645$ refers to an upper limit at 95\% confidence level. Besides the experimental uncertainties, a further uncertainty has to be considered, namely the uncertainty of the stopping power calculation entering Eq.~(\ref{equ:birks}). With the \textsc{srim} code, a determination of the stopping power for elemental materials is possible with an accuracy of a few percent \cite{zie99}, thus the described fitting procedure is repeated for  $dE/dx + \sigma_{dE/dx}$ and $dE/dx - \sigma_{dE/dx}$. Assuming a shift of 2\% of the stopping power already leads to a 2\% difference in the resulting quenching parameters. This uncertainty is added quadratically to the experimental one and is the dominant contribution to the resulting total uncertainty. The fit results are summarized in Tab.~\ref{tab:birksfit}. $C$ is consistent with zero for all investigated scintillator samples with an upper limit (95\%~CL) of about $10^{-7}$\,cm$^2$ MeV$^{-2}$ and the values for $kB$ range between (0.0094~$\pm$~0.0002)\,cm\,MeV$^{-1}$ and (0.0098~$\pm$~0.0003)\,cm\,MeV$^{-1}$. 

\setlength{\tabcolsep}{4pt}
\begin{table}[t]
\begin{center}
\caption{\label{tab:birksfit}  Quenching parameters $kB$ and $C$, following Eq.~(\ref{equ:birks}), determined within this work for different LAB based scintillators. The upper limits for $C$ are given for a confidence level of 95\%.}
\begin{tabular*}{0.48\textwidth}{lll}
\hline\noalign{\smallskip}
LAB admixture & $kB$ & $C$ \\
 & [cm MeV$^{-1}$] & [cm$^{2}$ MeV$^{-2}$] \\
\noalign{\smallskip}\hline\noalign{\smallskip}
2g/l\,PPO, 15mg/l\,bis-MSB & 0.0097\,$\pm$\,0.0002 & $\leq$\,5.0\,$\times 10^{-7}$\\
2g/l\,PPO & 0.0098\,$\pm$\,0.0003 & $\leq$\,4.0\,$\times 10^{-7}$ \\
3g/l\,PPO, 15mg/l\,bis-MSB & 0.0098\,$\pm$\,0.0003 & $\leq$\,1.0\,$\times 10^{-7}$\\
3g/l\,PPO & 0.0094\,$\pm$\,0.0002 & $\leq$\,6.5\,$\times 10^{-7}$\\
\noalign{\smallskip}\hline
\end{tabular*}
\end{center}
\end{table}

According to current knowledge, all ionization quen\-ching processes are primary processes in the scintillator, i.e. processes that transfer ionization energy to excitation energy of the solvent. Quenching processes compete with the excitation of the solvent molecules into $\pi$--electron singlet states \cite{bir64}. This is the primary process which is essential for the final scintillation emission in unitary systems as well as in binary and ternary systems. Consequently, for scintillators with the same solvent, the magnitude of ionization quenching should be the same independent of possible solutes. Within the presented measurement no significant deviation from these considerations could be observed.

For practical use, the proton quenching factor $Q_p$, the ratio of proton and electron light output functions

\begin{equation}
\label{equ:Qp}
Q_p(E) = \frac{L_p(E)}{L_e(E)},
\end{equation}

for each examined LAB scintillator is presented in Fig.~\ref{fig:qfact}. Above a kinetic energy of 8\,MeV, the proton light output is about half the electron light output. Below that energy the quenching factor strongly decreases, i.e. the proton light output is further reduced down to $<$20\% of the electron light output for $E\approx1$\,MeV.

\bildb{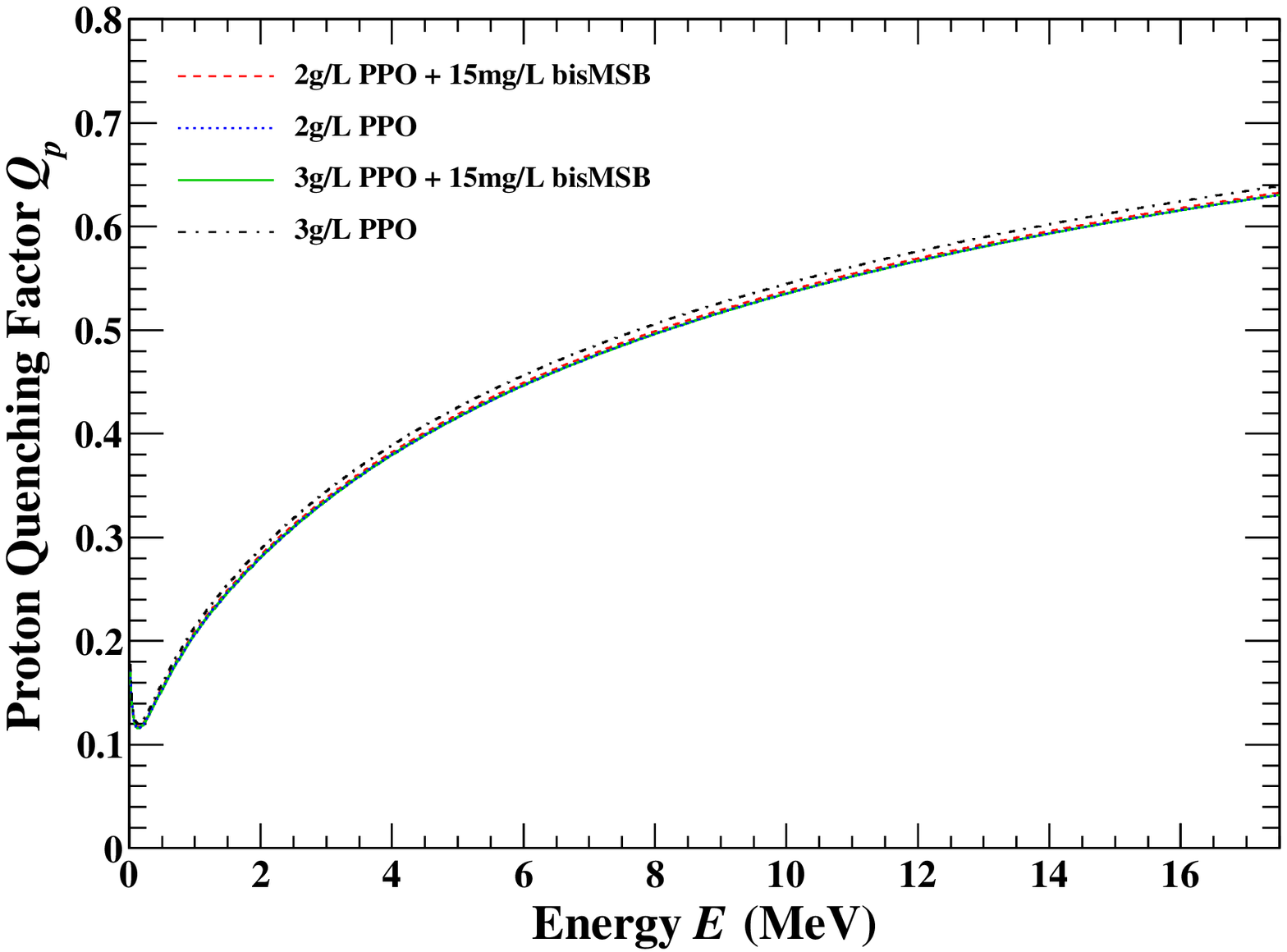}{0.5\textwidth}{Proton quenching factor $Q_p$ as function of kinetic energy $E$ for LAB based scintillators. The respective solute admixture is given in the legend.}{0.48\textwidth}{fig:qfact}{}{t}

\section{Application of the results}
\label{sec:appl}
The detection of supernova neutrinos of all flavors using neutrino--proton elastic scattering is  discussed in detail in \cite{bea11}. Within that reference, different experiments capable of the respective detection are compared, namely Borexino, KamLAND, SNO+ and LENA. For the LAB scintillator detector SNO+, a Birks' constant of $kB = 0.0073$\,cm\,MeV$^{-1}$ is used. However at that time no measurement of proton quenching in LAB was published. The measurements presented here for LAB with 2\,g/l~PPO (as will be used in SNO+) reveals with \mbox{$kB = 0.0098$\,cm\,MeV$^{-1}$} a stronger proton quenching. To show the impact of a stronger quenching on the event yield induced by a supernova, the same detector (i.e. SNO+) and assumptions for the supernova are used as in \cite{bea11}. In the same reference, the prospect of the larger detector LENA is discussed, assuming $kB = 0.01$\,cm\,MeV$^{-1}$ and $3.3\times10^{33}$ free protons in the liquid scintillator. As LENA favors LAB as solvent with 3\,g/l~PPO and 20\,mg/l~bis--MSB \footnote{M. Wurm (private communication).}, this detector is also discussed in the present article with the obtained result of $kB = 0.0098$\,cm\,MeV$^{-1}$ for LAB with 3\,g/l~PPO and 15\,mg/l~bis--MSB. For both detectors, $C$ is considered to be zero in agreement with the results summarized in Tab.~\ref{tab:birksfit}. 

The supernova is assumed to occur at a distance \mbox{$d=10$\,kpc}, releasing over the duration of the burst a total energy of $\varepsilon = 3\times10^{53}$\,erg which is equipartitioned among all flavors. The energy is distributed according to

\begin{eqnarray}
\label{equ:fluence}
	\frac{dF}{dE_{\nu}} &=& \sum \limits_{\alpha} \frac{dF_{\alpha}}{dE_{\nu}}\\
	&=& \frac{2.35 \times 10^{13}}{\mathrm{cm^2\,MeV}} \sum \limits_{\alpha} \frac{\varepsilon_{\alpha}}{d^2} \frac{E_{\nu}^3}{\langle E_{\nu}\rangle_{\alpha}^5} \mathrm{exp}\left(- \frac{4E_{\nu}}{\langle E_{\nu} \rangle_{\alpha}}\right), \nonumber
\end{eqnarray}

where $dF_{\alpha}/dE_{\nu}$ is the neutrino fluence per flavor. The energies are given in MeV, $d$ in 10\,kpc and $\varepsilon_{\alpha}$ in $10^{52}$\,erg. The mean energy $\langle E_{\nu} \rangle_{\alpha}$ of the individual flavors differs and is exemplarily set to 12\,MeV for $\nu_e$, 15\,MeV for \bnel and 18\,MeV for $\nu_x$, following the approach in \cite{bea11}. The respective fluence in $\nu_e$, \bnel and the sum of all \nx as well as the total fluence are shown in Fig.~\ref{fig:dFdEnu}. The fluences are of the order of $10^{10}$\,cm$^{-2}$\,MeV$^{-1}$ and the neutrino energy $E_{\nu}$ ranges up to about 60\,MeV.

\bildb{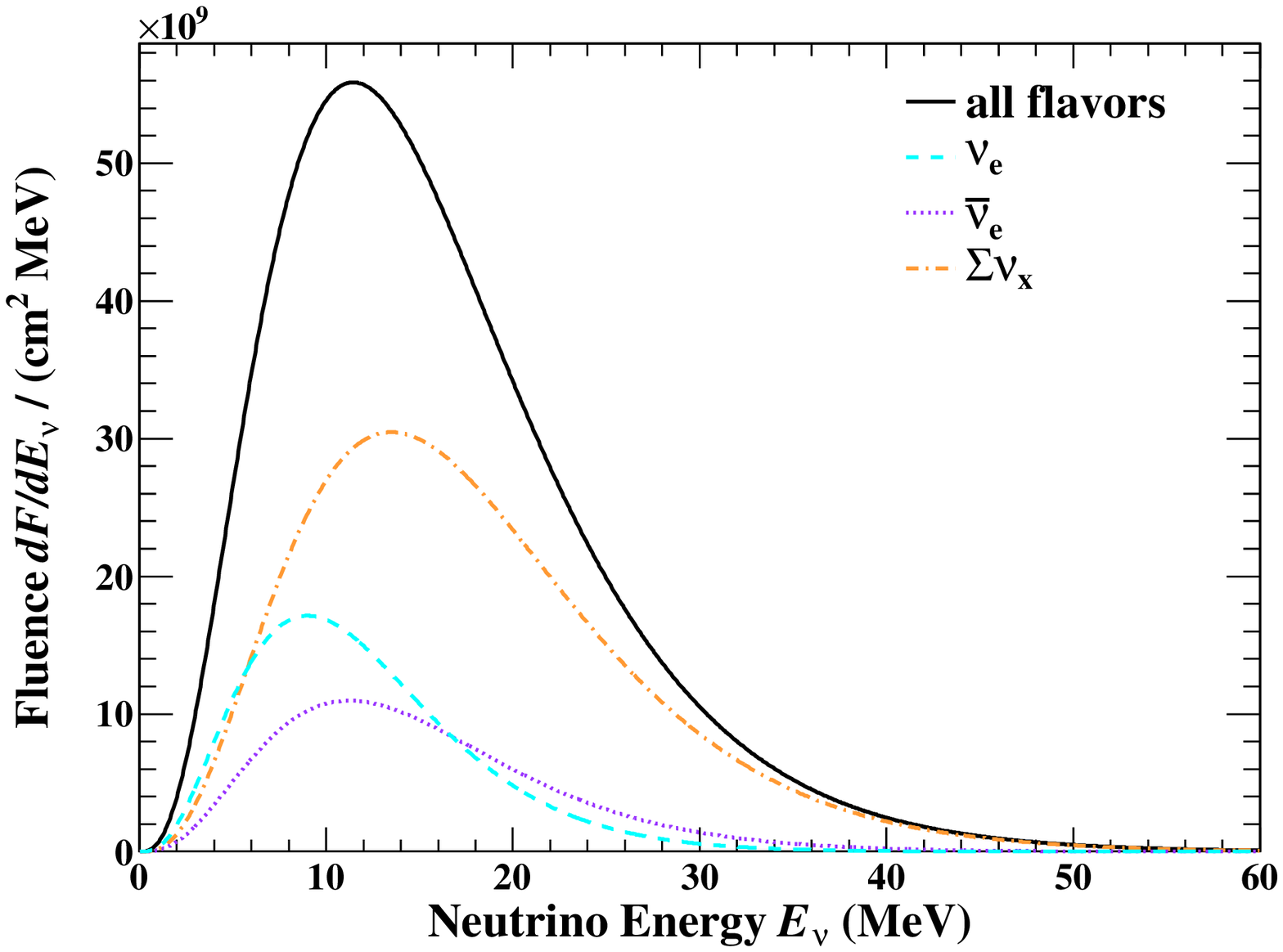}{0.5\textwidth}{Neutrino fluence on Earth induced by a 10\,kpc distant supernova, emitting a total energy of $3\times10^{53}$\,erg. The assumed mean neutrino energies are 12\,MeV for $\nu_e$, 15\,MeV for \bnel and 18\,MeV for $\nu_x$.}{0.48\textwidth}{fig:dFdEnu}{}{t}

For these low neutrino energies, the differential neu\-tri\-no--pro\-ton elastic scattering cross section \cite{wei72} can be simplified to

\begin{eqnarray}
\label{equ:xsec}
	\frac{d\sigma}{dE_{p}} &=& \frac{G_F^2 m_p}{\pi} \left[\left(1 - \frac{m_p E_p}{2 E_{\nu}^2} \right) c_V^2 + \left(1 - \frac{m_p E_p}{2 E_{\nu}^2} \right) c_A^2 \right]  \nonumber\\
	&\approx& \frac{4.83 \times 10^{-42} \mathrm{\,cm^2}}{\mathrm{MeV}} \left(1 + 466 \frac{E_p}{E_{\nu}^2}\right)
\end{eqnarray}

(considering only the lowest order in $E_{\nu}/m_{p}$) and  the cross section for neutrinos and anti--neutrinos can be regarded as identical \cite{bea02}. For the last step, a proton mass of $m_p = 938$\,MeV is used and the neutral--current coupling constants are  $c_V = 0.04$ and $c_A = 0.635$ \cite{bea02}. $E_{\nu}$ and the proton recoil energy $E_p$ are given in MeV. 

Using Eq.~(\ref{equ:fluence}) and Eq.~(\ref{equ:xsec}), the true recoil energy spectrum of recoil protons produced by supernova neutrinos is given by

\begin{equation}
\label{equ:dNdEp}
	\frac{dN}{dE_p} = N_p \int_{\epsilon_{\nu,min}}^{\infty} dE_{\nu} \frac{dF}{dE_{\nu}} \frac{d\sigma}{dE_{p}},
\end{equation}

where $N_p$ is the number of free protons in the scintillator target. The minimum neutrino energy needed to accelerate a proton to energy $E_p$ is $\epsilon_{\nu,min} \approx \sqrt{m_p E_p/2}$ \cite{bea02}. Fig.~\ref{fig:dNdEp} shows the expected proton recoil spectra for the various flavors in a detector with $N_p = 5.9 \times 10^{31}$ (corresponding to the SNO+ target mass of about 0.8\,kt \cite{loz12}). Despite the small cross section, the intense neutrino fluence apparently causes a significant neutrino--proton elastic scattering yield of the order of $10^2$ events per kt LAB. The proton energies range from 0\,MeV to about 5\,MeV.

\bildb{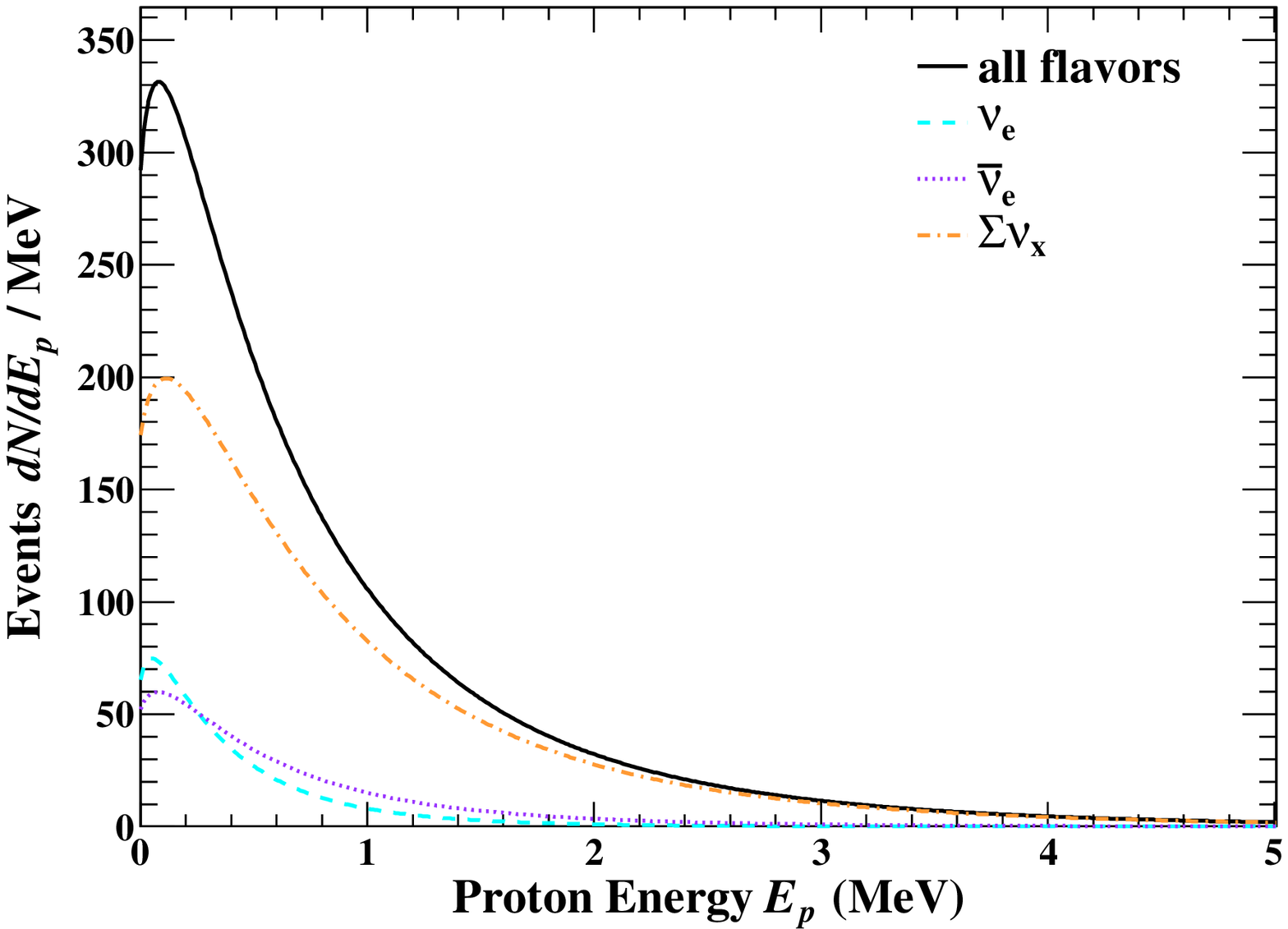}{0.5\textwidth}{True energy spectrum of recoiled protons in a $\sim$0.8\,kt LAB detector (like SNO+) from $\nu$--$p$ elastic scattering events, assuming the supernova neutrino fluences in Fig.~\ref{fig:dFdEnu}.}{0.48\textwidth}{fig:dNdEp}{}{t}

However, proton quenching markedly changes the shape of proton energy spectra and shifts the true proton recoil energies to lower observed energies. Even though there is no energy threshold to produce scintillation light, below 0.2\,MeV organic liquid scintillators are strongly affected by $\beta$--decays of  $^{14}$C, which is intrinsic to organic solvents. The high decay rate dominates this region and thus sets an effective threshold of 0.2\,MeV (further backgrounds are not taken into account within this article as they are described in detail in \cite{bea02} and found to be negligible for the presented supernova signal channel). Hence, the observed event yield finally depends on the quenching strength. The expected event yield above threshold will be discussed in the following for SNO+ and LENA. The respective detector specifications are given in Tab.~\ref{tab:SNdet}. The main difference between both detectors is the amount of target protons, which is about $5.9\times10^{31}$ in SNO+ and $3.25\times10^{33}$ in LENA, given a proton density in LAB of $6.31\times10^{22}$\,cm$^{-3}$ \cite{yeh08}.

\setlength{\tabcolsep}{7pt}
\begin{table}[t]
\caption{\label{tab:SNdet}  Detector properties assuming LAB with 2\,g/l~PPO as target in SNO+ and LAB with 3\,g/l~PPO and 15\,mg/l~bis--MSB in LENA. The SNO+ target mass and energy resolution is the same as employed in \cite{bea11} for better comparison of the results and in agreement with \cite{loz12}. The LENA target mass and resolution is taken from \cite{wur12}, where the expected resolution is quoted to be $\geq 200$\,p.e./MeV. The number of free protons $N_p$ is determined using a proton density in LAB of $6.31\times10^{22}$\,cm$^{-3}$ \cite{yeh08}. $C$ is assumed to be zero.}
\begin{tabular*}{0.48\textwidth}{lllll}
\hline\noalign{\smallskip}
Detector & Mass & $N_p$ & $E$ resolution & $kB$\\
 & [kt] & [$10^{31}$] & ($E$ in MeV) & [cm/MeV]\\
\noalign{\smallskip}\hline\noalign{\smallskip}
SNO+ & 0.8 & 5.9 & 5.0\%$/\sqrt{E}$ & 0.0098\\
LENA & 44 & 325 & 7.0\%$/\sqrt{E}$ & 0.0098\\
\noalign{\smallskip}\hline
\end{tabular*}
\end{table}

To convert the true proton energy spectra in Fig.~\ref{fig:dNdEp} into spectra of visible energy  $E^{\mathrm{vis}}_{p}$ in electron equivalent, the quenching factor $Q_p(E_p)$, as presented in Fig.~\ref{fig:qfact}, is applied to Eq.~(\ref{equ:dNdEp}). The resulting spectra are folded with the energy resolution and scaled according to the respective number of free protons as given in Tab.~\ref{tab:SNdet}. Fig.~\ref{fig:dNdEpq_SNOp} and Fig.~\ref{fig:dNdEpq_LENA} finally show the expected supernova neutrino signal spectra from $\nu$--$p$ elastic scattering in SNO+ and LENA under the given assumptions for the supernova scenario and the detectors. The event yield above a threshold of 0.2\,MeV, using $kB=0.0098$\,cm\,MeV$^{-1}$ and $C$ equal to zero for LAB, is about 98 events in SNO+ and 5403 events in LENA.  The yield scales linearly with the target size. For LAB scintillators it is in the range of $\sim$123 events/kt, which makes this detection channel very promising for large scale detectors. Applying the listed energy resolution only changes both yields by less than 0.5\% compared to perfect resolution and seems to have a negligible effect. The dependence on energy resolution thus is not further studied here, however, it needs to be investigated when a profound sensitivity study to supernova neutrino fluence parameters is performed. 

\bildb{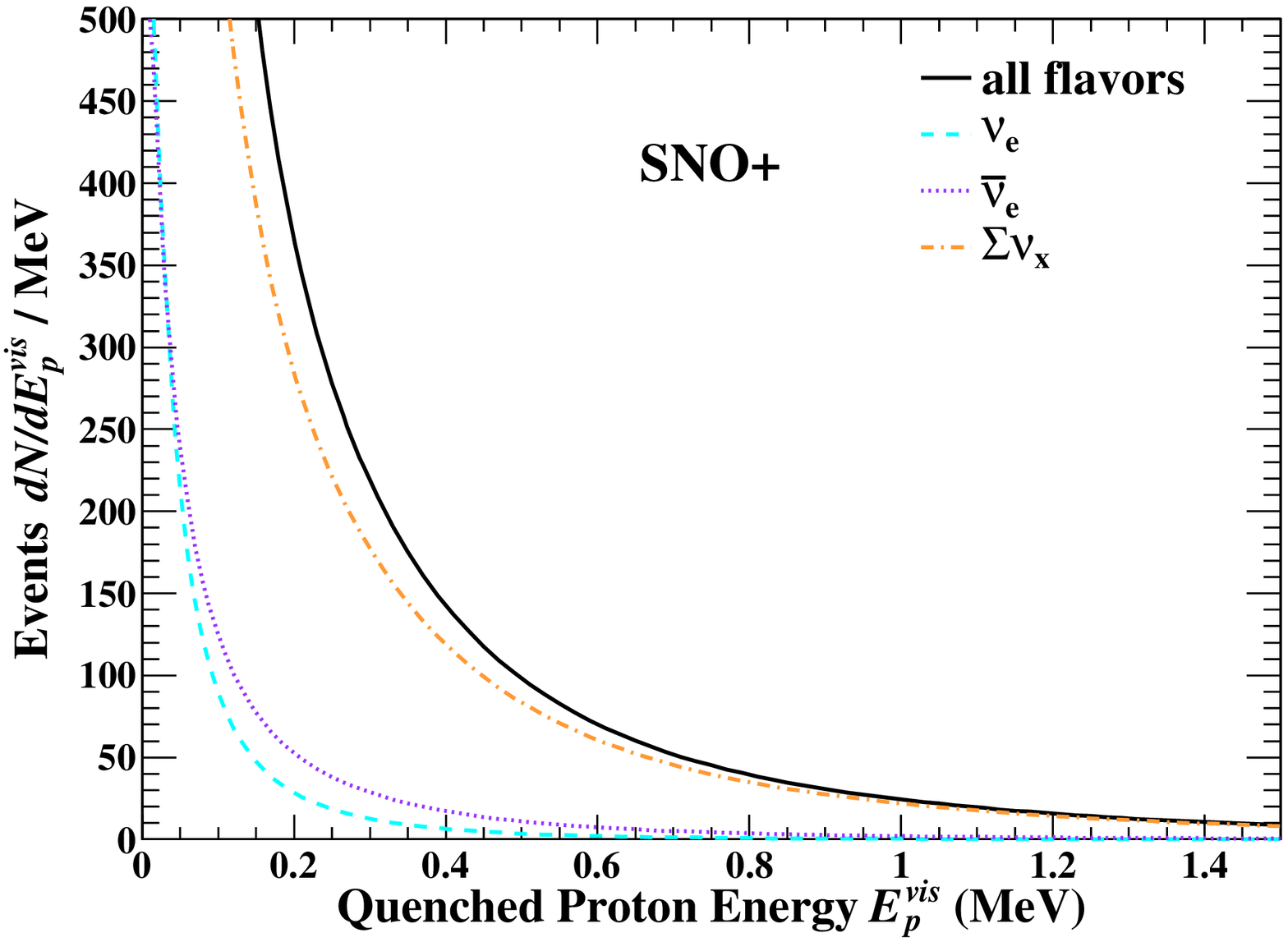}{0.5\textwidth}{Observed energy spectrum of recoiled protons in SNO+ considering proton quenching with \mbox{$kB=0.0098$\,cm\,MeV$^{-1}$} and an energy resolution of 5.0\%$/\sqrt{E}$. The supernova neutrino fluences in Fig.~\ref{fig:dFdEnu} are assumed.}{0.48\textwidth}{fig:dNdEpq_SNOp}{}{htbp}
\bildb{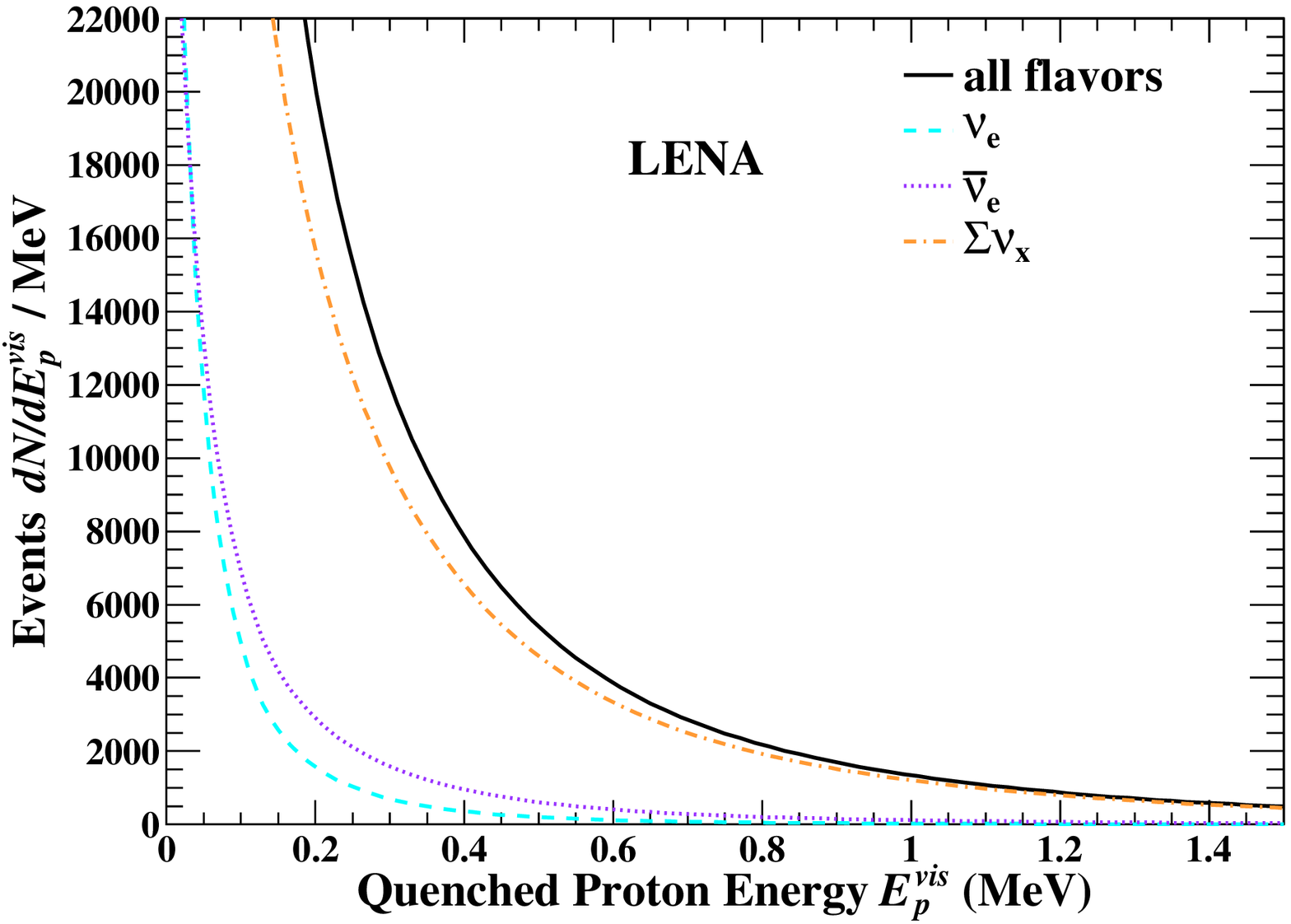}{0.5\textwidth}{Observed energy spectrum of recoiled protons in LENA considering proton quenching with \mbox{$kB=0.0098$\,cm\,MeV$^{-1}$} and an energy resolution of 7.0\%$/\sqrt{E}$. The supernova neutrino fluences in Fig.~\ref{fig:dFdEnu} are assumed.}{0.48\textwidth}{fig:dNdEpq_LENA}{}{htbp}

As mentioned earlier, the determination of $kB$ is most sensitive to the calculation of the stopping power $dE/dx$ and vice versa the calculation of $Q_p$ for a fixed Birks' constant. In reference \cite{bea11} $dE/dx$ tables from \textsc{pstar} \cite{pstar} are used while for the presented analysis \textsc{srim} tables are employed. Though the stopping power data is similar, the expected yield for SNO+ is recalculated with $kB=0.0073$\,cm\,MeV$^{-1}$ and $C$ considered to be zero, using \textsc{srim} tables, to show the impact of a higher $kB$ value without bias. Fig.~\ref{fig:dNdEp_all} compares the true proton energy sum spectrum and the observed sum spectra assuming $kB=0.0073$\,cm\,MeV$^{-1}$ and $kB=0.0098$\,cm\,MeV$^{-1}$ respectively. The higher quenching reduces the expected event yield by about 16\% from  $\sim$116 events\footnote{The yield quoted in \cite{bea11}, obtained with \textsc{pstar} tables, amounts to $\sim$111 events.} to $\sim$98 events. This figure demonstrates how proton quenching pushes supernova $\nu$--$p$ elastic scattering signal events to lower energies and that the yield above threshold strongly depends on the strength of the quenching. Therefore it is of great importance for liquid scintillator detectors, sensitive to supernova neutrinos, to properly measure the proton quenching parameters of the used scintillator to extract valuable information about supernova neutrino fluences of all flavors.

\bildb{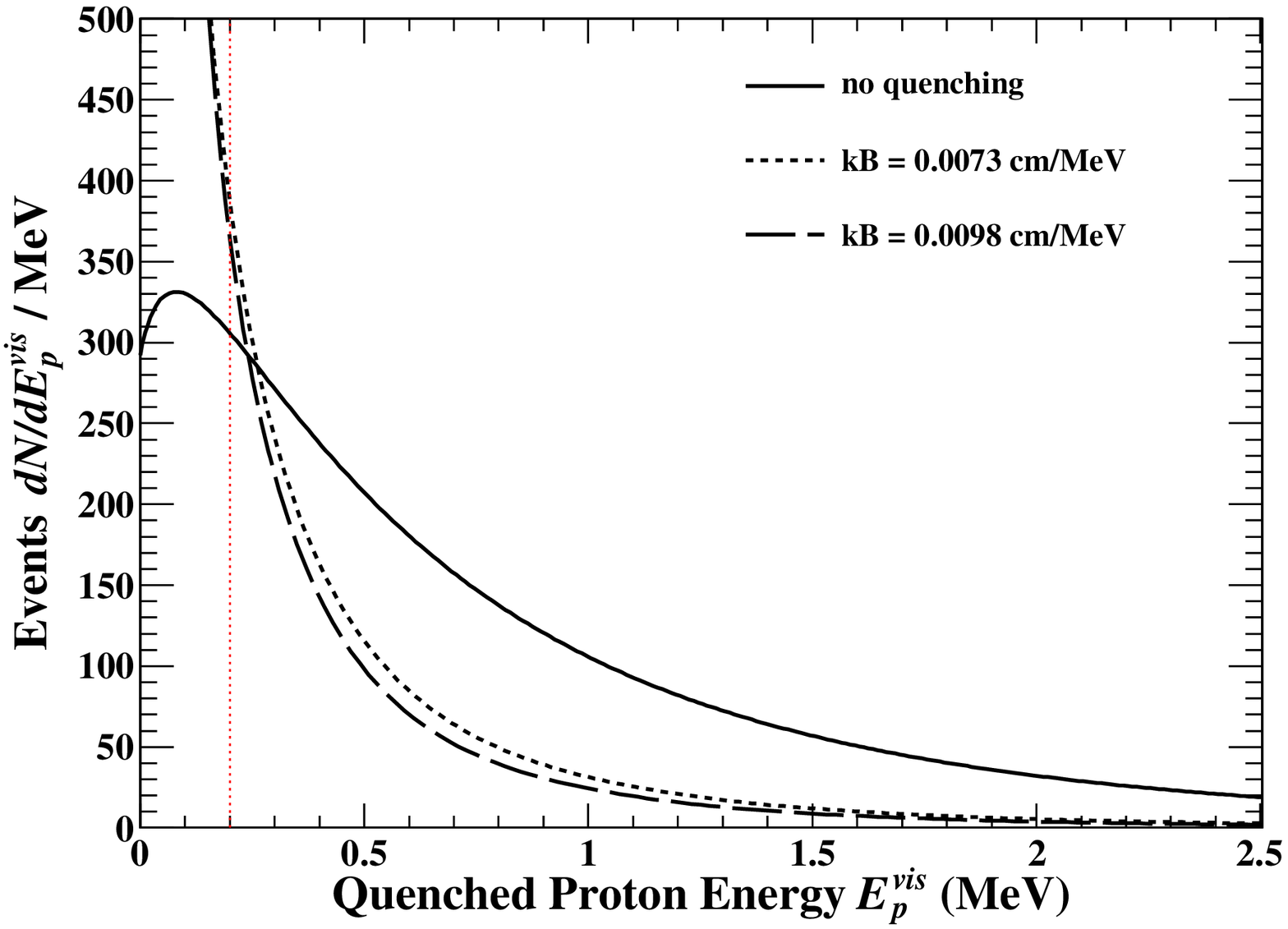}{0.5\textwidth}{Observed energy spectrum of recoiled protons in SNO+ considering an energy resolution of 5.0\%$/\sqrt{E}$ and a supernova neutrino fluence as in Fig.~\ref{fig:dFdEnu}. Shown is the sum event spectrum for three different proton quenching scenarios, i.e. no quenching, $kB=0.0073$\,cm\,MeV$^{-1}$ and \mbox{$kB=0.0098$\,cm\,MeV$^{-1}$}. The dotted line indicates a threshold of 200\,keV.}{0.48\textwidth}{fig:dNdEp_all}{}{t}

\section{Summary and outlook}
\label{sec:sum}
This article presents a measurement of the light output function $L(E)$ of protons relative to the one of electrons for four different LAB based scintillators. A measurement of the relative proton light output reveals the magnitude of light output reduction due to ionization quenching. Fitting the functional description of $L(E)$ proposed by J.\,B.\,Birks, parameterized by $kB$ and $C$, to the experimental data yielded values for $kB$ ranging between (0.0094~$\pm$~0.0002)\,cm\,MeV$^{-1}$ and (0.0098 $\pm$ 0.0003)\,cm\,MeV$^{-1}$ and an upper limit of about $10^{-7}$\,cm$^2$ MeV$^{-2}$ (95\%~CL) for the parameter $C$ for the investigated scintillators. 

It should be noted, however, that the parameters within Birks' law, $kB$ and $C$, alone are not conclusive for the comparison of different scintillators (i.e. scintillators with different solvents) as they are correlated with the stopping power of the scintillator material. Consequently, to decide between two different scintillators, always the full light output functions need to be compared. Different solutes in the same solvent, on the other hand, do not show a difference in the calculated stopping powers due to their small concentration. For the latter case, $kB$ and $C$ are thus sensitive to a potential influence on the strength of ionization quenching by different solutes. The presented measurement does not reveal a significant disparity of the resulting parameters of all LAB based scintillators and is in accordance with the expectation that all ionization quenching processes are primary processes.

A further remark is that the values for $kB$ extracted from measurements of the non--linearity in the light output of low--energy electrons is not necessarily identical to the one of heavy ionizing particles like ions \cite{tre12}. Whereas it is assumed that  the $kB$ value is the same for different ions, like protons and alphas, once the conditions of the measurement are fixed, as discussed in detail in \cite{tre10}. This hypothesis can be tested with the taken data by determining the $\alpha$ light output function for the investigated scintillators. $^{12}$C(n,$\alpha$)$^9$Be and $^{12}$C(n,n'\,3$\alpha$) reactions, induced by neutrons from the beam, lead to characteristic structures in the pulse--height response of the scintillator which can be reproduced with the \textsc{nresp7} code. A proper knowledge of the scintillator response to $\alpha$'s is important for the development of a sophisticated background model of a liquid scintillator detector, including $\alpha$--induced backgrounds. This analysis and discussion though is beyond the scope of this article and an outlook to future possibilities.

Using the measured proton light output functions for LAB solutions, the potential of LAB based scintillator detectors to observe supernova neutrinos of all flavors via $\nu$--$p$ elastic scattering is discussed and it is shown how a different Birks constant affects the event yield. Due to this impact on the observation, well--known quenching parameters are important for liquid scintillator detectors, capable of supernova neutrino detection.

\begin{small}
\begin{acknowledgements}
We thank Minfang Yeh and Torben Ferber for helpful discussions and comments on the manu\-script. We also thank the mechanical workshop of the TU~Dresden for the production of the scintillator cell as well as Kai Tittelmeier and the accelerator staff of the PTB for their support. The LAB solvent was provided by Petresa~Canada~Inc., B\'{e}cancour. This work was supported by the Deutsche Forschungsgemeinschaft (DFG).
\end{acknowledgements}
\end{small}



\end{document}